\documentclass[journal,10pt,twocolumn,twoside]{IEEEtran}
\usepackage{multicol}   
\usepackage{multirow}
\usepackage[pdftex]{graphicx}
\usepackage{algorithm}
\usepackage{algpseudocode}
\usepackage[small]{caption}
\usepackage{float}
\usepackage{amssymb,amsmath}
\usepackage{graphicx}
\usepackage{subfigure}
\usepackage{xspace}
\usepackage{amsthm}
\usepackage{fancyhdr}

\usepackage{caption,setspace}
\usepackage{amssymb,amsmath}
\usepackage{bm}
\usepackage{textcomp} 
\usepackage{epstopdf}
\usepackage{bbding}
\usepackage{cite} 
\usepackage{color}
\usepackage{pifont}
\usepackage{lipsum}
\usepackage{amstext}
\usepackage{orcidlink}

\usepackage{cuted}
\usepackage{stfloats}

\graphicspath{{fig/}}

\newcommand\scalemath[2]{\scalebox{#1}{\mbox{\ensuremath{\displaystyle #2}}}}

\begin{document}

\title{\huge Exploiting Movable-Element STARS for Rate Splitting Multiple Access}
\author{Muhammad Asif \,\orcidlink{0000-0002-9699-1675}, \IEEEmembership{Member, IEEE}, Asim Ihsan \,\orcidlink{0000-0001-7491-7178}, Irfan Muhammad \,\orcidlink{0000-0001-5703-7988}, \IEEEmembership{Member, IEEE}, Mohd Hamza Naim Shaikh  \,\orcidlink{0000-0002-1869-0009}, \IEEEmembership{Member, IEEE}, Muhammad Ayzed Mirza \,\orcidlink{0000-0003-3176-2764}, \IEEEmembership{Member, IEEE}, Zhu Shoujin \,\orcidlink{0000-0002-2615-2489}, and Symeon Chatzinotas \,\orcidlink{0000-0001-5122-0001}, \IEEEmembership{Fellow, IEEE} 

\thanks{This work was supported in part by the Tongling University Talent Research Initiation Fund Project under Grant No. 2023tlxyrc17; and in part by the Anhui Provincial Higher Education Institutions Young Core Faculty Domestic Visiting Scholar Training Funding Program under Grant No. JNFX2025067.}  

\thanks{Muhammad Asif and Zhu Shoujin are with the School of Electrical and Information Engineering, Tongling University, Tongling 244002, China (e-mails: masif@tlu.edu.cn, 2023028@tlu.edu.cn).
	
 Asim Ihsan is with the Interdisciplinary Research Center for Communication Systems and Sensing, King Fahd University of Petroleum \& Minerals (KFUPM ) Dhahran, Saudi Arabia (e-mail: asim.ihsan@kfupm.edu.sa).
	
 Irfan Muhammad is with the Centre for Wireless Communications, University of Oulu, 90570 Oulu, Finland (e-mail: irfan.muhammad@oulu.fi).
  
M. H. N. Shaikh  is with the School of Electronics, Electrical Engineering and Computer Science (EEECS), Queen's University Belfast, Belfast BT9 5BN, U.K (e-mail: h.shaikh@qub.ac.uk).

Muhammad Ayzed Mirza is with  School of Computer Science and Information Engineering, Internet of Vehicles Lab, Qilu Institute of Technology, Jinan, Shandong 250200 P.R, China (e-mail: ayzed.mirza@gmail.com ).

 Symeon Chatzinotas is with the Interdisciplinary Centre for Security, Reliability and Trust (SnT), University of Luxembourg, 1855 Luxembourg City, Luxembourg (e-mail: symeon.chatzinotas@uni.lu).

}

\vspace{-0.7cm}}%

\markboth{}
{ \MakeLowercase{\textit{}}} 
\maketitle

\begin{abstract}
\begin{abstract}
  This paper investigates a movable-element simultaneously transmitting and reflecting reconfigurable intelligent surface (ME-STARS) assisted rate-splitting multiple access (RSMA) system under imperfect channel state information (CSI). Unlike conventional STARS with fixed element positions, the elements of ME-STARS can be repositioned within a predefined region, providing additional spatial degrees of freedom for improving the cascaded transmitter--STARS--user channels. To exploit this flexibility while accounting for CSI uncertainty, we formulate a robust sum-rate maximization problem that jointly optimizes the transmit beamforming, common-rate allocation, reflection and transmission coefficients, and ME-STARS element positions, subject to transmit-power, user-rate, minimum inter-element spacing, and movement-region constraints. The resulting problem is highly non-convex due to the strong coupling among the design variables and the position-dependent channels. To address this challenge, an iterative optimization framework is developed in which the transmit beamforming, STARS coefficients, and element positions are successively optimized through tractable convex reformulations. In particular, the element positions are updated sequentially using a majorization--minimization (MM) framework, where quadratic surrogate functions are constructed from the first- and second-order derivatives of the position-dependent channels while preserving the minimum inter-element spacing constraint. Simulation results demonstrate that the proposed ME-STARS design consistently outperforms the considered benchmark schemes. Moreover, the performance gains remain significant under increasing CSI uncertainty, highlighting the effectiveness of element repositioning for robust RSMA transmission.
\end{abstract}

\end{abstract}


\IEEEpeerreviewmaketitle


\section{Introduction}
\IEEEPARstart{T} {he} growing demand for high-rate transmission and massive connectivity in future wireless networks has placed increasing emphasis on improving spectral efficiency and making more effective use of the available spatial resources \cite{wang2022gcwcn,nguyen20216g}. Multiple-input multiple-output (MIMO) technology can provide substantial spatial multiplexing and beamforming gains, but the large number of radio-frequency chains required by conventional multi-antenna systems may lead to considerable hardware cost and energy consumption \cite{wang2025optimal,busari2017millimeter}. Reconfigurable intelligent surface (RIS) technology has therefore attracted significant attention as a cost- and energy-efficient approach for reshaping wireless propagation through programmable passive elements \cite{liu2021reconfigurable,asif2025noma}. Nevertheless, the reflection-only operation of conventional RIS restricts user coverage to one side of the surface. To overcome this limitation, simultaneously transmitting and reflecting surface (STARS) has emerged as a more flexible architecture capable of serving users located in both the transmission and reflection regions \cite{mu2021simultaneously,asif2026robust}. In parallel, rate-splitting multiple access (RSMA) has gained increasing attention as an efficient multiuser transmission strategy by partially decoding interference and partially treating it as noise, thereby providing greater flexibility in interference management \cite{mao2022rate}. The integration of STARS and RSMA therefore offers a powerful means of combining full-space coverage with flexible interference management, thereby improving spectral efficiency and multiuser transmission performance \cite{asif2024leveraging}. Motivated by the complementary advantages of STARS and RSMA, several recent studies have explored their integration in diverse scenarios, including integrated sensing and communications (ISAC) \cite{liu2025star}, intelligent transportation systems \cite{zhang2025star}, covert communications \cite{zhang2024covert}, autonomous aerial vehicles \cite{perdana2026energy}, and physical-layer security \cite{wang2025maximizing}, demonstrating the versatility of STARS-RSMA integration across a wide range of communication applications.

Despite these advantages, conventional RIS and STARS architectures generally rely on elements deployed at fixed positions, which limits the available spatial degrees of freedom and restricts their ability to adapt
to changes in the wireless propagation environment\cite{liu2025star, zhang2025star,perdana2026energy,wang2025maximizing}. Although deploying more elements can provide additional array gain, it also increases channel estimation overhead, implementation complexity, and circuit power consumption \cite{wang2024reconfigurable}. This tradeoff motivates the development of more flexible architectures that can exploit the available spatial region more efficiently, rather than relying solely on a larger number of fixed-position elements. Inspired by recent advances in position-adjustable antenna technologies such as fluid antennas (FAs) \cite{wong2021fluid} and movable antennas (MAs) \cite{zhu2023modeling}, antenna positions can be flexibly adjusted within a predefined region to exploit spatial channel variations and establish more favorable channel conditions. This idea has also been extended to movable-element RIS (ME-RIS) architectures \cite{hokmabadi2026joint, zhao2026movable,zhou2025movable,hu2024intelligent,hokmabadi2026secure}, where the reflecting elements can be repositioned within a predefined region to provide greater flexibility in controlling the propagation environment and shaping the cascaded channels. For instance, the work in \cite{hokmabadi2026joint} investigated an ME-RIS-assisted full-duplex MISO system and jointly optimized the element positions and beamforming design to maximize the achievable sum rate. In \cite{zhao2026movable}, an ME-RIS-assisted communication system was investigated, where the positions of the movable elements were optimized to maximize the achievable communication rate. The work in \cite{zhou2025movable} considered an ME-RIS-assisted communication system and maximized the achievable rate by jointly optimizing the element positions and reflection coefficients of the ME-RIS. The authors in \cite{hu2024intelligent} proposed a ME-RIS to eliminate phase-distribution mismatch under Rician fading, enabling a unified non-uniform phase-shift design with lower complexity and improved performance. The work in \cite{hokmabadi2026secure} investigated secure energy-efficient full-duplex transmission with MAs and ME-RIS and developed a hybrid meta-learning framework to jointly optimize beamforming, power allocation, and movable-element positions.

Inspired by the benefits offered by ME-RIS, extending element mobility to STARS provides additional flexibility, since repositioning the elements can simultaneously affect the transmission and reflection
links and reshape the corresponding cascaded channels. This has motivated the development of movable-element STARS (ME-STARS) \cite{zhao2026exploiting,zhao2025movable,zhu2025movable}, where the element positions can be jointly optimized with the reflection and transmission coefficients to further improve multiuser communication performance. The authors in \cite{zhao2026exploiting} proposed an ME-STARS-assisted multiuser communication framework and jointly optimized the transmit beamforming, reflection/transmission coefficients, and element positions for different STARS operating protocols. The results showed that ME-STARS achieves notable weighted sum-rate gains over fixed-position STARS (FP-STARS) and ME-RIS counterparts. In \cite{zhao2025movable}, the authors investigated an ME-STARS-assisted secure communication system and jointly optimized the active beamforming, passive beamforming, and ME positions to maximize the sum secrecy rate. The work in \cite{zhu2025movable} proposed an ME-STARS-assisted near-field wideband communication framework, where the STARS element positions and beamforming are jointly optimized to mitigate the beam-squint effect. The results showed that ME-STARS can effectively improve near-field wideband performance while providing a favorable tradeoff between performance gain and hardware complexity.

Building upon the above discussion, the integration of ME-STARS and RSMA offers a promising opportunity to jointly exploit spatial reconfigurability and flexible interference management in multiuser communications. In particular, ME-STARS can reshape the cascaded channels by adjusting the element positions and reflection/transmission coefficients, while RSMA can efficiently manage inter-user interference through common and private message transmission. However, the existing literature still exhibits the following limitations: \textbf{1)} Existing works on ME-STARS have mainly focused on conventional multiuser transmission \cite{zhao2026exploiting}, physical-layer security \cite{zhao2025movable}, and near-field wideband communications \cite{zhu2025movable}, while the integration of ME-STARS and RSMA remains largely unexplored. Therefore, the potential gains arising from the joint exploitation of element mobility, reflection/transmission control, and rate splitting have not yet been fully exploited; \textbf{2)} Moreover, most existing ME-STARS studies rely on the assumption of perfect channel state information (CSI), which may be difficult to guarantee in practical systems. This issue becomes more critical in an ME-STARS-assisted RSMA system because CSI uncertainty affects not only the transmit beamforming and reflection/transmission coefficients, but also the common-rate allocation and element-position optimization. Since the element positions directly determine the cascaded channel responses, inaccurate CSI may lead to inappropriate position adjustments and consequently reduce the spatial gains offered by ME-STARS. Furthermore, the coupling among the position-dependent channels, RSMA transmission, and reflection/transmission design makes robust resource allocation considerably more challenging. 

To the best of our knowledge, a robust transmission design for ME-STARS-assisted RSMA systems that jointly optimizes the transmit beamforming, common-rate allocation, reflection/transmission coefficients, and movable-element positions under CSI uncertainty has not yet been investigated. Motivated by the aforementioned research gaps, the main contributions of this work are summarized as follows.
\begin{itemize}
\item We develop a robust ME-STARS-assisted RSMA framework for multi-user communications under imperfect CSI. By jointly exploiting the movable element positions and reflection/transmission coefficients, the proposed
design introduces additional spatial flexibility for shaping the cascaded channels while RSMA provides flexible interference management. Based on this framework, a robust sum-rate maximization problem is formulated by jointly optimizing the transmit beamforming, common-rate allocation, reflection/transmission coefficients, and ME-STARS element positions, subject to the movement-region and minimum inter-element spacing
constraints.

\item To address the strong coupling among the design variables and the non-convexity of the formulated problem, an iterative optimization framework is developed by successively handling the transmit beamforming,
reflection/transmission coefficient, and ME-STARS element-position optimization subproblems. In particular, the movable elements are updated sequentially using a majorization--minimization (MM) framework, where
quadratic surrogate functions are constructed from the first- and second-order derivatives of the position-dependent channels while satisfying the movement-region and minimum inter-element spacing
constraints.

\item Different from existing ME-STARS designs that largely rely on perfect CSI, the proposed framework explicitly accounts for CSI uncertainty in the joint design of RSMA transmission, ME-STARS reflection/transmission coefficients, and movable-element positions. Since the movable-element positions directly determine the cascaded channel responses, the resulting robust design captures the coupled impact of CSI uncertainty on transmit beamforming, common-rate allocation, reflection/transmission coefficients,
and element-position optimization.

\item Finally, extensive numerical results demonstrate that the proposed robust ME-STARS-assisted RSMA design consistently outperforms the considered benchmark schemes. Moreover, the proposed algorithm exhibits
stable and fast convergence under the considered system configurations.

\end{itemize}  
            
 \begin{figure}[!t]
	\centering
	\includegraphics [width=0.50\textwidth]{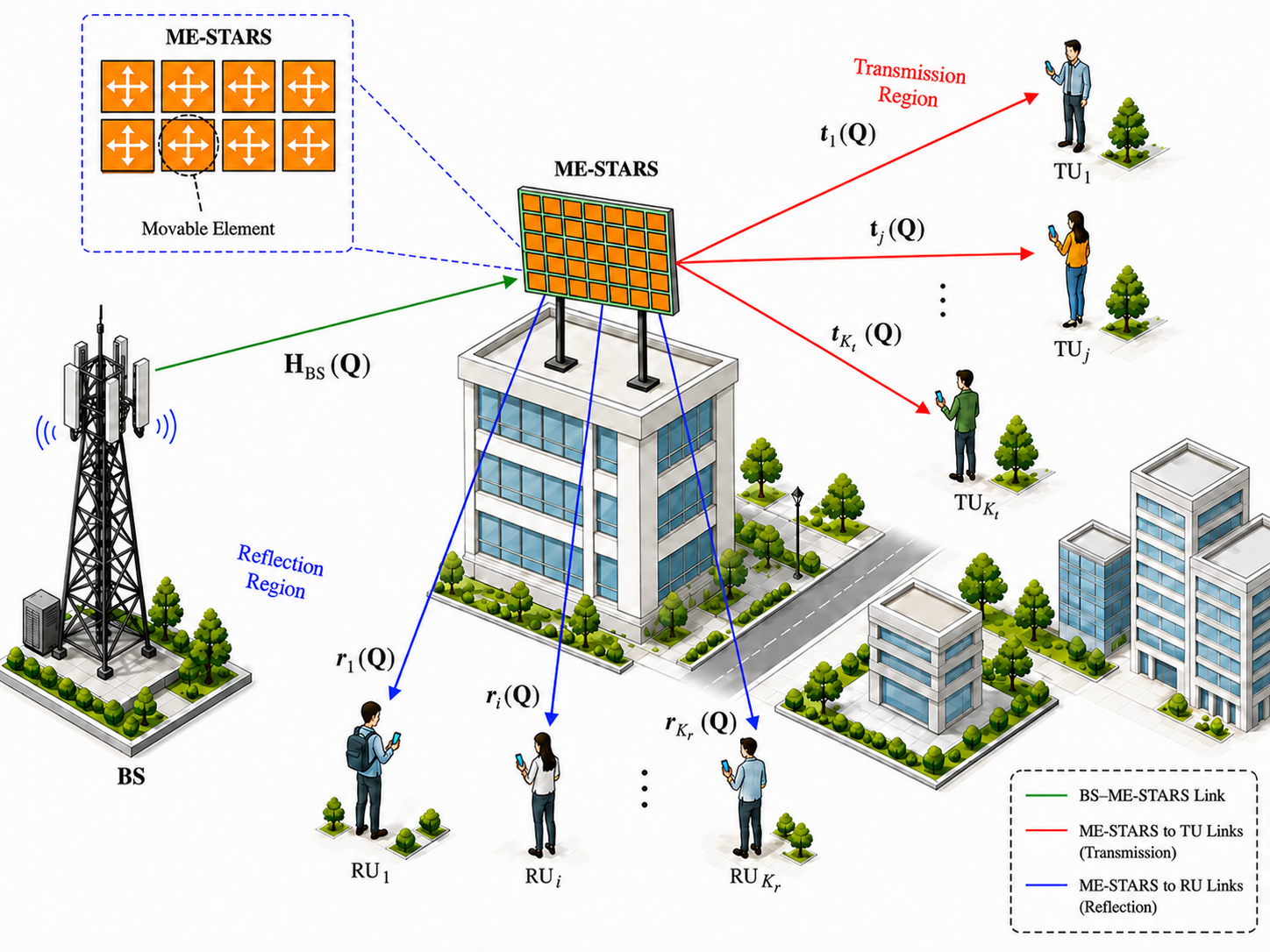}
	\caption{Illustration of system model.}
	\label{f1}
\end{figure} 
            
\section{System Model and Problem Formulation}
\subsection{ME-STARS-Assisted RSMA System Model}

We consider a downlink RSMA system assisted by a ME-STARS, as illustrated in Fig.~\ref{f1}. The BS is equipped with $N_{\rm B}$ antennas and serves multiple single-antenna users through an ME-STARS comprising $N_{\rm S}$ movable elements. The reflection- and transmission-region user index sets are denoted by
$\mathcal K_{\rm r}=\{1,\ldots,K_{\rm r}\}$ and
$\mathcal K_{\rm t}=\{1,\ldots,K_{\rm t}\}$, respectively. To avoid ambiguity between the two region-specific index sets, define the disjoint tagged user sets $\mathcal U_{\rm r}\triangleq\{({\rm r},i):i\in\mathcal K_{\rm r}\}, \ \mathcal U_{\rm t}\triangleq\{({\rm t},j):j\in\mathcal K_{\rm t}\}, \ \mathcal U\triangleq\mathcal U_{\rm r}\cup\mathcal U_{\rm t}$. The $N_{\rm S}$ elements of the ME-STARS are movable and can be repositioned within a predefined square region $\mathcal C\subset\mathbb R^2$ of size $A\times A$. The
position of the $n$-th movable element is denoted by $\mathbf q_n=[q_{x,n},q_{y,n}]^T\in\mathbb R^2$, while the
overall element configuration is represented by $\mathbf Q=[\mathbf q_1,\mathbf q_2,\ldots,\mathbf q_{N_{\rm S}}]\in\mathbb R^{2\times N_{\rm S}}$. Accordingly, the position of each movable element satisfies
$\mathbf q_n\in\mathcal A$, $\forall n=1,\ldots,N_{\rm S}$.

Furthermore, to maintain a physically feasible element configuration and mitigate mutual coupling, the minimum
inter-element spacing constraint is imposed as
\begin{equation}
	\|\mathbf q_n-\mathbf q_m\|_2\geq d_{\min},
	\qquad
	\forall n\neq m,
	\label{eq:minimum_spacing}
\end{equation}
where $d_{\min}$ denotes the prescribed minimum allowable distance between any two distinct movable elements.

Additionally, the reflection and transmission response matrices of ME-STARS are expressed as
\begin{align}
	\boldsymbol{\Xi}_{\rm r}
	&=
	\operatorname{diag}
	\left(
	\rho_1e^{j\omega_1},\ldots,
	\rho_{N_{\rm S}}e^{j\omega_{N_{\rm S}}}
	\right),
	\label{eq:Xi_r}\\
	\boldsymbol{\Xi}_{\rm t}
	&=
	\operatorname{diag}
	\left(
	\tau_1e^{j\varphi_1},\ldots,
	\tau_{N_{\rm S}}e^{j\varphi_{N_{\rm S}}}
	\right),
	\label{eq:Xi_t}
\end{align}
where $\rho_n$ and $\tau_n$ are the reflection and transmission amplitudes, whereas $\omega_n$ and $\varphi_n$ are the corresponding phase shifts. 

Further, we adopt the energy-splitting (ES) protocol for the ME-STARS \cite{asif2023energy}, under which the incident power at each element is partitioned into reflected and transmitted components. Accordingly, the corresponding coefficients satisfy
\begin{align}
	\rho_n^2+\tau_n^2&\leq1,
	&
	0\leq\rho_n,\tau_n&\leq1,
	&
	0\leq\omega_n,\varphi_n&<2\pi,
	\qquad \forall n.
	\label{eq:surface_constraints}
\end{align}

Next, the position-dependent steering vector of the ME-STARS is expressed as \cite{zhu2023modeling}:
\begin{equation}
	\mathbf u(\vartheta,\varpi,\mathbf Q)
	=
	\left[
	e^{jk_0\Delta_1(\vartheta,\varpi,\mathbf q_1)},
	\ldots,
	e^{jk_0\Delta_{N_{\rm S}}
		(\vartheta,\varpi,\mathbf q_{N_{\rm S}})}
	\right]^T,
	\label{eq:ME_STARS_steering}
\end{equation}
where $\vartheta$ and $\varpi$ denote the azimuth and elevation angles of the corresponding propagation direction, respectively, $k_0=2\pi/\lambda$ is the wavenumber, and
$\lambda$ denotes the carrier wavelength. Moreover, $\Delta_n(\vartheta,\varpi,\mathbf q_n)=q_{x,n}\sin(\vartheta)\cos(\varpi)+q_{y,n}\sin(\varpi)$ denotes the propagation path difference introduced by the position of the $n$-th movable element relative to the reference point of the ME-STARS. Consequently, varying $\mathbf q_n$ changes the phase response of the corresponding channel component.

Assuming a half-wavelength-spaced uniform linear array at the BS, its steering vector in the direction of $\vartheta_{\rm out}$ is given as
\begin{equation}
	\mathbf b(\vartheta_{\rm out})
	=
	\left[
	1,e^{j\pi\sin(\vartheta_{\rm out})},\ldots,
	e^{j(N_{\rm B}-1)\pi\sin(\vartheta_{\rm out})}
	\right]^T.
	\label{eq:BS_steering}
\end{equation}

Further, under the far-field Rician steering-response channel model, the BS-to-ME-STARS channel is given by
\begin{align}
	\mathbf H_{\rm BS}(\mathbf Q)
	&=
	\sqrt{\ell_{\rm BS}}
	\left(
	\sqrt{\frac{\chi_{\rm BS}}{\chi_{\rm BS}+1}}
	\mathbf u(\vartheta_{\rm in},\varpi_{\rm in},\mathbf Q)
	\mathbf b^H(\vartheta_{\rm out})
	\right.
	\nonumber\\[-1mm]
	&\hspace{28mm}\left.
	+
	\sqrt{\frac{1}{\chi_{\rm BS}+1}}
	\mathbf Z_{\rm BS}
	\right),
	\label{eq:H_BS}
\end{align}
where $\ell_{\rm BS}$ and $\chi_{\rm BS}$ denote the large-scale channel gain and Rician factor, respectively. The angles $\vartheta_{\rm in}$ and $\varpi_{\rm in}$ specify the arrival direction at the ME-STARS, whereas $\vartheta_{\rm out}$ denotes the departure direction from the BS. Moreover, $\mathbf Z_{\rm BS}\in\mathbb C^{N_{\rm S}\times N_{\rm B}}$ represents the non-line-of-sight (NLoS) component of the BS--ME-STARS channel, whose entries are independently distributed as $\mathcal{CN}(0,1)$. The BS--ME-STARS link spans a relatively long propagation distance and may encounter environmental scattering; therefore, both the LoS and NLoS components are retained through the Rician channel model.

Additionally, the ME-STARS is assumed to be deployed such that the ME-STARS--user links have unobstructed propagation and are comparatively shorter than the BS--ME-STARS link. Hence, the LoS components are
assumed to dominate the scattered components, and the corresponding ME-STARS--user channels are modeled as \cite{liu2026joint}:
\begin{align}
	\mathbf r_i^H(\mathbf Q)
	&=
	\sqrt{\ell_{{\rm r},i}}
	\mathbf u^H(\vartheta_{{\rm r},i},\varpi_{{\rm r},i},\mathbf Q),
	&& i\in\mathcal K_{\rm r},
	\label{eq:r_channel}\\
	\mathbf t_j^H(\mathbf Q)
	&=
	\sqrt{\ell_{{\rm t},j}}
	\mathbf u^H(\vartheta_{{\rm t},j},\varpi_{{\rm t},j},\mathbf Q),
	&& j\in\mathcal K_{\rm t}.
	\label{eq:t_channel}
\end{align}

Subsequently, for a generic tagged user $u\in\mathcal U$, we define $\mathbf h_{{\rm s},u}(\mathbf Q)=\mathbf r_i(\mathbf Q)$ and $\boldsymbol{\Xi}_u=\boldsymbol{\Xi}_{\rm r}$ for $u=({\rm r},i)$, whereas $\mathbf h_{{\rm s},u}(\mathbf Q)=\mathbf t_j(\mathbf Q)$ and $\boldsymbol{\Xi}_u=\boldsymbol{\Xi}_{\rm t}$ for $u=({\rm t},j)$. Accordingly, the estimated cascaded channel is
\begin{equation}
	\widehat{\mathbf g}_u(\mathbf Q)
	=
	\mathbf h_{{\rm s},u}^H(\mathbf Q)
	\boldsymbol{\Xi}_u^H
	\mathbf H_{\rm BS}(\mathbf Q),
	\qquad u\in\mathcal U.
	\label{eq:equivalent_channel}
\end{equation}

Let $s_0$ denote the common stream decoded by all users, and let
$s_u$ denote the private stream intended for user $u$. The BS transmit signal is
\begin{equation}
	\mathbf x
	=
	\mathbf v_0s_0
	+
	\sum_{u\in\mathcal U}\mathbf v_us_u,
	\label{eq:tx_signal}
\end{equation}
where $\mathbf v_0$ and $\mathbf v_u$ are the corresponding beamforming vectors. The common stream $s_0$ and the private streams $\{s_u\}_{u\in\mathcal U}$ are assumed to be mutually independent and normalized to unit power, i.e., $\mathbb E\{|s_0|^2\}=1$ and $\mathbb E\{|s_u|^2\}=1$, $\forall u\in\mathcal U$. Accordingly, the total transmit power at the BS is constrained by
\begin{equation}
	\|\mathbf v_0\|_2^2
	+
	\sum_{u\in\mathcal U}\|\mathbf v_u\|_2^2
	\leq P_{\max},
	\label{eq:power_constraint}
\end{equation}
where $P_{\max}$ denotes the maximum available transmit power at the BS.

Thus, the received signal of user $u$ is expressed as
\begin{align}
	y_u
	&=
	\mathbf g_u(\mathbf Q)\mathbf v_0s_0
	+
	\sum_{v\in\mathcal U}
	\mathbf g_u(\mathbf Q)\mathbf v_vs_v
	+
	n_u,
	\label{eq:received_signal}.
\end{align}
where $n_u\sim\mathcal{CN}(0,\sigma_u^2)$.

\subsection{Channel Uncertainty and Robust Achievable Rates}
In practice, the equivalent cascaded channels cannot be perfectly estimated. For each user $u\in\mathcal U$, let $\mathbf g_u(\mathbf Q)\in\mathbb C^{1\times N_{\rm B}}$ and $\widehat{\mathbf g}_u(\mathbf Q)\in\mathbb C^{1\times N_{\rm B}}$ denote the actual and estimated equivalent cascaded channels, respectively. Accordingly, the corresponding channel outer-product matrices are defined as $\mathbf G_u(\mathbf Q)\triangleq\mathbf g_u^H(\mathbf Q)\mathbf g_u(\mathbf Q)\in\mathbb C^{N_{\rm B}\times N_{\rm B}}$, $\widehat{\mathbf G}_u(\mathbf Q)\triangleq\widehat{\mathbf g}_u^H(\mathbf Q)\widehat{\mathbf g}_u(\mathbf Q)\in\mathbb C^{N_{\rm B}\times N_{\rm B}}$. Then, for tractable robust design, the channel estimation error is modeled using the following spectral-norm-bounded uncertainty set \cite{li2022robust,zhang2023robust}:
\begin{align}
	\mathbf G_u(\mathbf Q)
	&=
	\widehat{\mathbf G}_u(\mathbf Q)
	+
	\Delta\mathbf G_u,
	\qquad
	\mathbf G_u(\mathbf Q)\succeq\mathbf 0,
	\label{eq:Gram_uncertainty}\\
	\Delta\mathbf G_u
	&=
	\Delta\mathbf G_u^H,
	\qquad
	\|\Delta\mathbf G_u\|_2\leq\eta_u,
	\qquad
	u\in\mathcal U,
	\nonumber
\end{align}
where $\Delta\mathbf G_u$ represents the channel outer-product uncertainty matrix, and $\eta_u\geq 0$ denotes its uncertainty radius.

To facilitate the subsequent robust reformulation, we introduce the lifted transmit covariance matrices $\mathbf V_0\triangleq\mathbf v_0\mathbf v_0^H$ and $\mathbf V_u\triangleq\mathbf v_u\mathbf v_u^H$,
$u\in\mathcal U$. Moreover, let $\mathbf V_{\rm p}\triangleq \sum_{j\in\mathcal U}\mathbf V_j$ and $\mathbf V_{-u}\triangleq\mathbf V_{\rm p}-\mathbf V_u$, $u\in\mathcal U$. Here, $\mathbf V_{\rm p}$ denotes the combined covariance matrix of all private streams, whereas $\mathbf V_{-u}$ represents the aggregate covariance matrix of the private streams intended for all users except user $u$. For any positive semidefinite transmit covariance matrix
$\mathbf V\succeq\mathbf 0$, the duality between the spectral
and nuclear norms yields \cite{asif2026robust,zhang2023robust}:
\begin{align}
	\left|
	\operatorname{Tr}
	\left(
	\Delta\mathbf G_u\mathbf V
	\right)
	\right|
	&\leq
	\|\Delta\mathbf G_u\|_2
	\|\mathbf V\|_*\leq
	\eta_u\operatorname{Tr}(\mathbf V),
	\label{eq:dual_norm_bound}
\end{align}
where $\|\mathbf V\|_*$ denotes the nuclear norm of
$\mathbf V$, defined as the sum of its singular values. Since
$\mathbf V\succeq\mathbf 0$, its nuclear norm satisfies
$\|\mathbf V\|_*=\operatorname{Tr}(\mathbf V)$. Accordingly,
the uncertain received-power term is bounded as
\begin{align}
	\operatorname{Tr}
	\left[
	\left(
	\widehat{\mathbf G}_u
	-
	\eta_u\mathbf I_{N_{\rm B}}
	\right)
	\mathbf V
	\right]
	&\leq
	\operatorname{Tr}
	\left(
	\mathbf G_u\mathbf V
	\right)
	\nonumber\\
	&\leq
	\operatorname{Tr}
	\left[
	\left(
	\widehat{\mathbf G}_u
	+
	\eta_u\mathbf I_{N_{\rm B}}
	\right)
	\mathbf V
	\right].
	\label{eq:robust_trace_bound}
\end{align}

When decoding the common stream, all private streams are treated as interference. Accordingly, the worst-case common-stream SINR at user $u$ is given by
\begin{equation}
	\overline{\Gamma}_u^0
	=
	\frac{
		\operatorname{Tr}
		\left[
		\left(
		\widehat{\mathbf G}_u-\eta_u\mathbf I_{N_{\rm B}}
		\right)\mathbf V_0
		\right]
	}{
		\operatorname{Tr}
		\left[
		\left(
		\widehat{\mathbf G}_u+\eta_u\mathbf I_{N_{\rm B}}
		\right)\mathbf V_{\rm p}
		\right]
		+
		\sigma_u^2
	},
	\qquad u\in\mathcal U.
	\label{eq:robust_common_SINR}
\end{equation}
After removing the common stream through successive interference cancellation, user $u$ decodes its private stream while treating the remaining private streams as interference. Its worst-case private-stream SINR is
\begin{equation}
	\overline{\Gamma}_u^{\rm p}
	=
	\frac{
		\operatorname{Tr}
		\left[
		\left(
		\widehat{\mathbf G}_u-\eta_u\mathbf I_{N_{\rm B}}
		\right)\mathbf V_u
		\right]
	}{
		\operatorname{Tr}
		\left[
		\left(
		\widehat{\mathbf G}_u+\eta_u\mathbf I_{N_{\rm B}}
		\right)\mathbf V_{-u}
		\right]
		+
		\sigma_u^2
	},
	\qquad u\in\mathcal U.
	\label{eq:robust_private_SINR}
\end{equation}
The corresponding robust rates are
\begin{equation}
	\overline R_u^0
	=
	\log_2(1+\overline\Gamma_u^0),
	\qquad
	\overline R_u^{\rm p}
	=
	\log_2(1+\overline\Gamma_u^{\rm p}).
	\label{eq:robust_rates}
\end{equation}

Let $c_u\geq 0$ denote the portion of the common rate allocated
to user $u$, and define
$\mathbf c\triangleq[c_u]_{u\in\mathcal U}$. The overall
common rate is given by
\begin{equation}
	R_0(\mathbf c)
	\triangleq
	\sum_{u\in\mathcal U}c_u.
	\label{eq:common_rate_allocation}
\end{equation}
Since the common stream must be decoded by every user, the
overall common rate cannot exceed the common-stream decoding
rate achievable at any user. Therefore,
\begin{equation}
	R_0(\mathbf c)
	\leq
	\overline R_u^0,
	\qquad
	\forall u\in\mathcal U.
	\label{eq:common_decoding}
\end{equation}
Accordingly, the robust achievable rate of user $u$ is
expressed as
\begin{equation}
	\overline R_u
	=
	c_u+\overline R_u^{\rm p},
	\qquad
	\forall u\in\mathcal U.
	\label{eq:total_user_rate}
\end{equation}

\subsection{Problem Formulation}
We formulate the robust sum-rate maximization problem by jointly optimizing the BS beamforming vectors, the common-rate allocation, the reflection and transmission coefficients of the ME-STARS, and the positions of its movable elements, as follows:
\begin{subequations}
	\label{eq:joint_problem}
	\begin{align}
		\max_{\mathcal X}\quad
		&
		\sum_{u\in\mathcal U}\overline R_u
		\label{eq:joint_objective}\\
		\mathrm{s.t.}\quad
		&
		R_0(\mathbf c)\leq\overline R_u^0,
		\qquad \forall u\in\mathcal U,
		\label{eq:joint_common}\\
		&
		\overline R_u\geq R_u^{\min},
		\qquad \forall u\in\mathcal U,
		\label{eq:joint_QoS}
		\end{align}
		\begin{align}
		&
		\|\mathbf v_0\|_2^2
		+
		\sum_{u\in\mathcal U}\|\mathbf v_u\|_2^2
		\leq P_{\max},
		\label{eq:joint_power}\\
		&
		\mathbf q_n\in\mathcal C,
		\qquad \forall n,
		\label{eq:joint_region}\\
		&
		\|\mathbf q_n-\mathbf q_m\|_2\geq d_{\min},
		\qquad \forall n\neq m,
		\label{eq:joint_spacing}\\
		&\rho_n^2+\tau_n^2\leq1,
		\qquad \forall n,
		\label{eq:joint_energy}\\
		&0\leq\rho_n,\tau_n\leq1,
		\qquad \forall n,
		\label{eq:joint_amplitude}\\
		&0\leq\omega_n,\varphi_n<2\pi,
		\qquad \forall n,
		\label{eq:joint_phase}\\
		&
		c_u\geq0,
		\qquad \forall u\in\mathcal U.
		\label{eq:joint_common_nonnegative}
	\end{align}
\end{subequations}
where $\mathcal X=\left\{\mathbf v_0,\{\mathbf v_u\}_{u\in\mathcal U},\boldsymbol{\Xi}_{\rm r}, \boldsymbol{\Xi}_{\rm t},\mathbf Q,\mathbf c\right\}$. Problem \eqref{eq:joint_problem} is non-convex because the beamforming vectors, the ME-STARS coefficients, and the positions of its movable elements are coupled in the robust rate expressions. In addition, the position-dependent steering responses and the minimum-spacing constraints are non-convex. We therefore employ an alternating-optimization strategy in which the individual variable blocks are updated successively.

\section{Proposed Optimization Framework}

\subsection{Active Beamforming Optimization}
For fixed ME-STARS element positions $\mathbf Q$ and fixed
reflection/transmission coefficients, define $\mathcal X_{\rm A}\triangleq\left\{\mathbf v_0,\{\mathbf v_u\}_{u\in\mathcal U},\mathbf c\right\}$.
The active beamforming subproblem is formulated as
\begin{subequations}
	\label{eq:active_beamforming_problem}
	\begin{align}
		\max_{\mathcal X_{\rm A}}\quad
		&
		\sum_{u\in\mathcal U}\overline R_u
		\label{eq:active_beamforming_objective}\\
		\mathrm{s.t.}\quad
		&
		R_0(\mathbf c)
		\leq
		\overline R_u^0,
		\qquad
		\forall u\in\mathcal U,
		\label{eq:active_beamforming_common}\\
		&
		\overline R_u
		\geq
		R_u^{\min},
		\qquad
		\forall u\in\mathcal U,
		\label{eq:active_beamforming_QoS}\\
		&
		\|\mathbf v_0\|_2^2
		+
		\sum_{u\in\mathcal U}
		\|\mathbf v_u\|_2^2
		\leq
		P_{\max},
		\label{eq:active_beamforming_power}\\
		&
		c_u\geq0,
		\qquad
		\forall u\in\mathcal U.
		\label{eq:active_beamforming_common_nonnegative}
	\end{align}
\end{subequations}

Further, for notational convenience, we define $\mathbf G_u^{-}\triangleq\widehat{\mathbf G}_u-\eta_u\mathbf I, \ \mathbf G_u^{+}\triangleq \widehat{\mathbf G}_u+\eta_u\mathbf I$. Then, the robust common- and private-stream power terms can be rewritten as
\begin{align}
	A_u^0(\mathcal V)
	&=
	\operatorname{Tr}
	\left(
	\mathbf G_u^{-}\mathbf V_0
	\right)
	+
	\operatorname{Tr}
	\left(
	\mathbf G_u^{+}\mathbf V_{\rm p}
	\right)
	+
	\sigma_u^2,
	\label{eq:active_A0}\\
	B_u^0(\mathcal V)
	&=
	\operatorname{Tr}
	\left(
	\mathbf G_u^{+}\mathbf V_{\rm p}
	\right)
	+
	\sigma_u^2,
	\label{eq:active_B0}\\
	A_u^{\rm p}(\mathcal V)
	&=
	\operatorname{Tr}
	\left(
	\mathbf G_u^{-}\mathbf V_u
	\right)
	+
	\operatorname{Tr}
	\left(
	\mathbf G_u^{+}\mathbf V_{-u}
	\right)
	+
	\sigma_u^2,
	\label{eq:active_Ap}\\
	B_u^{\rm p}(\mathcal V)
	&=
	\operatorname{Tr}
	\left(
	\mathbf G_u^{+}\mathbf V_{-u}
	\right)
	+
	\sigma_u^2,
	\label{eq:active_Bp}
\end{align}
where $\mathcal V \triangleq\left\{\mathbf V_0,\{\mathbf V_u\}_{u\in\mathcal U}\right\}$.

Accordingly, the robust common- and private-stream rates are
\begin{align}
	\overline R_u^0(\mathcal V)
	&=
	\log_2 A_u^0(\mathcal V)
	-
	\log_2 B_u^0(\mathcal V),
	\label{eq:active_common_rate}\\
	\overline R_u^{\rm p}(\mathcal V)
	&=
	\log_2 A_u^{\rm p}(\mathcal V)
	-
	\log_2 B_u^{\rm p}(\mathcal V).
	\label{eq:active_private_rate}
\end{align}

Since $\mathbf V_0=\mathbf v_0\mathbf v_0^H,\ \mathbf V_u=\mathbf v_u\mathbf v_u^H$, the equivalent
covariance-matrix formulation of the active beamforming problem is
given by
\begin{subequations}
	\label{eq:exact_active_problem}
	\begin{align}
		\max_{\mathcal V,\mathbf c}\quad
		&
		R_0(\mathbf c)
		+
		\sum_{u\in\mathcal U}
		\overline R_u^{\rm p}(\mathcal V)
		\label{eq:exact_active_objective}\\
		\mathrm{s.t.}\quad
		&
		R_0(\mathbf c)
		\leq
		\overline R_u^0(\mathcal V),
		\qquad
		\forall u\in\mathcal U,
		\label{eq:exact_active_common}
				\end{align}
		\begin{align}
		&
		c_u+
		\overline R_u^{\rm p}(\mathcal V)
		\geq
		R_u^{\min},
		\qquad
		\forall u\in\mathcal U,
		\label{eq:exact_active_QoS}\\
		&
		\operatorname{Tr}(\mathbf V_0)
		+
		\operatorname{Tr}(\mathbf V_{\rm p})
		\leq
		P_{\max},
		\label{eq:exact_active_power}\\
		&
		\mathbf V_0\succeq\mathbf 0,
		\qquad
		\mathbf V_u\succeq\mathbf 0,
		\quad
		\forall u\in\mathcal U,
		\label{eq:exact_active_PSD}\\
		&
		\operatorname{rank}(\mathbf V_0)\leq1,
		\qquad
		\operatorname{rank}(\mathbf V_u)\leq1,
		\quad
		\forall u\in\mathcal U,
		\label{eq:exact_active_rank}\\
		&
		c_u\geq0,
		\qquad
		\forall u\in\mathcal U.
		\label{eq:exact_active_common_nonnegative}
	\end{align}
\end{subequations}
Problem \eqref{eq:exact_active_problem} is non-convex due to the difference-of-logarithm rate expressions and the rank-one constraints.

We first employ semidefinite relaxation (SDR) by relaxing the rank-one constraints in \eqref{eq:exact_active_rank}. To facilitate the convex approximation of the non-convex rate expressions, we
introduce the positive slack variables
$\alpha_u^0$, $\beta_u^0$, $\alpha_u^{\rm p}$, and
$\beta_u^{\rm p}$, which satisfy
$\alpha_u^0\leq A_u^0(\mathcal V)$,
$B_u^0(\mathcal V)\leq\beta_u^0$,
$\alpha_u^{\rm p}\leq A_u^{\rm p}(\mathcal V)$, and
$B_u^{\rm p}(\mathcal V)\leq\beta_u^{\rm p}$,
$\forall u\in\mathcal U$.

Next, at iteration $\ell$, we set $\beta_u^{0,(\ell)}\triangleq B_u^0\bigl(\mathcal V^{(\ell)}\bigr)$ and
$\beta_u^{{\rm p},(\ell)}\triangleq B_u^{\rm p}\bigl(\mathcal V^{(\ell)}\bigr)$,
$\forall u\in\mathcal U$.
By exploiting the concavity of $\log_2(x)$, its first-order
Taylor expansion at $x^{(\ell)}>0$ provides the following
global affine upper approximation:
\begin{equation}
	\log_2(x)
	\leq
	\log_2\bigl(x^{(\ell)}\bigr)
	+
	\frac{x-x^{(\ell)}}
	{x^{(\ell)}\ln 2},
	\qquad x>0.
	\label{eq:active_log_upper_generic}
\end{equation}

Accordingly, the logarithmic interference terms $\log_2(\beta_u^0)$ and $\log_2(\beta_u^{\rm p})$ can be upper-bounded as
\begin{align}
	\log_2\bigl(\beta_u^0\bigr)
	&\leq
	\log_2\bigl(\beta_u^{0,(\ell)}\bigr)
	+
	\frac{
		\beta_u^0-\beta_u^{0,(\ell)}
	}{
		\beta_u^{0,(\ell)}\ln 2
	} \qquad u\in\mathcal U,
	\label{eq:active_log_B0_upper}\\
	\log_2\bigl(\beta_u^{\rm p}\bigr)
	&\leq
	\log_2\bigl(\beta_u^{{\rm p},(\ell)}\bigr)
	+
	\frac{
		\beta_u^{\rm p}-\beta_u^{{\rm p},(\ell)}
	}{
		\beta_u^{{\rm p},(\ell)}\ln 2
	},
	\qquad u\in\mathcal U.
	\label{eq:active_log_Bp_upper}
\end{align}

Therefore, concave lower bounds of the robust rates are given as
\begin{align}
	\widetilde R_u^{0,(\ell)}
	&=
	\log_2\alpha_u^0
	-
	\log_2\beta_u^{0,(\ell)}
	-
	\frac{
		\beta_u^0-\beta_u^{0,(\ell)}
	}{
		\beta_u^{0,(\ell)}\ln2
	},
	\label{eq:active_common_rate_lower}\\
	\widetilde R_u^{{\rm p},(\ell)}
	&=
	\log_2\alpha_u^{\rm p}
	-
	\log_2\beta_u^{{\rm p},(\ell)}
	-
	\frac{
		\beta_u^{\rm p}-\beta_u^{{\rm p},(\ell)}
	}{
		\beta_u^{{\rm p},(\ell)}\ln2
	}.
	\label{eq:active_private_rate_lower}
\end{align}

 We further introduce the auxiliary rate variables $r_u^0$ and $r_u^{\rm p}$ for the common and private streams, respectively. Accordingly, the resulting active beamforming subproblem is reformulated as
\begin{subequations}
	\label{eq:convex_active_problem}
	\begin{align}
		\max_{\mathcal Y_{\rm A}}\quad
		&
		R_0(\mathbf c)
		+
		\sum_{u\in\mathcal U}r_u^{\rm p}
		\label{eq:convex_active_objective}\\
		\mathrm{s.t.}\quad
		&
		R_0(\mathbf c)
		\leq
		r_u^0,
		\qquad
		\forall u\in\mathcal U,
		\label{eq:convex_active_common}\\
		&
		c_u+r_u^{\rm p}
		\geq
		R_u^{\min},
		\qquad
		\forall u\in\mathcal U,
		\label{eq:convex_active_QoS}\\
		&
		r_u^0
		\leq
		\widetilde R_u^{0,(\ell)},
		\qquad
		\forall u\in\mathcal U,
		\label{eq:convex_active_common_rate}\\
		&
		r_u^{\rm p}
		\leq
		\widetilde R_u^{{\rm p},(\ell)},
		\qquad
		\forall u\in\mathcal U,
		\label{eq:convex_active_private_rate}\\
		&
		\alpha_u^0
		\leq
		A_u^0(\mathcal V),
		\qquad
		\forall u\in\mathcal U,
		\label{eq:convex_active_A0}\\
		&
		B_u^0(\mathcal V)
		\leq
		\beta_u^0,
		\qquad
		\forall u\in\mathcal U,
		\label{eq:convex_active_B0}\\
		&
		\alpha_u^{\rm p}
		\leq
		A_u^{\rm p}(\mathcal V),
		\qquad
		\forall u\in\mathcal U,
		\label{eq:convex_active_Ap}
					\end{align}
		\begin{align}
		&
		B_u^{\rm p}(\mathcal V)
		\leq
		\beta_u^{\rm p},
		\qquad
		\forall u\in\mathcal U,
		\label{eq:convex_active_Bp}\\
		&
		\alpha_u^0\geq\xi,\quad
		\beta_u^0\geq\xi,\quad
		\alpha_u^{\rm p}\geq\xi,\quad
		\beta_u^{\rm p}\geq\xi,
		\qquad
		\forall u\in\mathcal U,
		\label{eq:convex_active_positive}\\
		&
		\operatorname{Tr}(\mathbf V_0)
		+
		\operatorname{Tr}(\mathbf V_{\rm p})
		\leq
		P_{\max},
		\label{eq:convex_active_power}\\
		&
		\mathbf V_0\succeq\mathbf 0,
		\qquad
		\mathbf V_u\succeq\mathbf 0,
		\quad
		\forall u\in\mathcal U,
		\label{eq:convex_active_PSD}\\
		&
		c_u\geq0,
		\qquad
		\forall u\in\mathcal U,
		\label{eq:convex_active_common_nonnegative}
	\end{align}
\end{subequations}
where $\mathcal Y_{\rm A}\triangleq\big\{\mathbf V_0,\{\mathbf V_u\}_{u\in\mathcal U},\mathbf c, \{r_u^0,r_u^{\rm p}\}_{u\in\mathcal U},\{\alpha_u^0,\beta_u^0,\alpha_u^{\rm p}\\,\beta_u^{\rm p}\}_{u\in\mathcal U}\big\}$. Problem \eqref{eq:convex_active_problem} is a convex semidefinite program (SDP), since the approximated rate expressions in \eqref{eq:active_common_rate_lower} and \eqref{eq:active_private_rate_lower} are concave, while the remaining constraints are either affine or semidefinite. Therefore, it can be efficiently solved using standard convex optimization solvers. Upon convergence, if the obtained covariance matrices satisfy the rank-one conditions, the corresponding beamforming vectors are recovered through eigenvalue decomposition. Otherwise, Gaussian randomization is applied to obtain feasible rank-one beamforming solutions \cite{ni2021resource}.

\subsection{ME-STARS Passive Beamforming Optimization}
For fixed active beamforming vectors and ME-STARS element positions $\mathbf Q$, we define $\mathcal X_{\rm P}\triangleq \left\{\{\rho_n,\tau_n,\omega_n,\varphi_n\}_{n=1}^{N_{\rm S}},\mathbf c\right\}$ as the set of optimization variables associated with the passive beamforming design. Accordingly, the ME-STARS passive beamforming subproblem is formulated as
\begin{subequations}
	\label{eq:passive_beamforming_problem}
	\begin{align}
		\max_{\mathcal X_{\rm P}}\quad
		&
		\sum_{u\in\mathcal U}\overline R_u
		\label{eq:passive_beamforming_objective}\\
		\mathrm{s.t.}\quad
		&
		R_0(\mathbf c)
		\leq
		\overline R_u^0,
		\qquad
		\forall u\in\mathcal U,
		\label{eq:passive_beamforming_common}\\
		&
		\overline R_u
		\geq
		R_u^{\min},
		\qquad
		\forall u\in\mathcal U,
		\label{eq:passive_beamforming_QoS}\\
		&
		\rho_n^2+\tau_n^2
		\leq
		1,
		\qquad
		\forall n,
		\label{eq:passive_beamforming_energy}\\
		&
		0\leq\rho_n,\tau_n\leq1,
		\qquad
		\forall n,
		\label{eq:passive_beamforming_amplitude}\\
		&
		0\leq\omega_n,\varphi_n<2\pi,
		\qquad
		\forall n,
		\label{eq:passive_beamforming_phase}\\
		&
		c_u\geq0,
		\qquad
		\forall u\in\mathcal U.
		\label{eq:passive_beamforming_common_nonnegative}
	\end{align}
\end{subequations}

The passive beamforming design jointly optimizes the reflection and transmission coefficients of the ME-STARS together with the common-rate allocation vector $\mathbf c$. To this end, the reflection and transmission coefficient vectors are expressed as
\begin{align}
	\boldsymbol{\theta}_{\rm r}
	&\triangleq
	\left[
	\rho_1 e^{j\omega_1},
	\ldots,
	\rho_{N_{\rm S}}e^{j\omega_{N_{\rm S}}}
	\right]^T,
	\nonumber\\
	\boldsymbol{\theta}_{\rm t}
	&\triangleq
	\left[
	\tau_1 e^{j\varphi_1},
	\ldots,
	\tau_{N_{\rm S}}e^{j\varphi_{N_{\rm S}}}
	\right]^T.
	\label{eq:passive_vectors}
\end{align}

For each user $u\in\mathcal U$, define
\begin{equation}
	x_u
	\triangleq
	\begin{cases}
		{\rm r}, & u\in\mathcal U_{\rm r},\\
		{\rm t}, & u\in\mathcal U_{\rm t},
	\end{cases}
	\qquad
	\boldsymbol{\Xi}_u
	=
	\operatorname{diag}
	\left(
	\boldsymbol{\theta}_{x_u}
	\right).
	\label{eq:user_region_index}
\end{equation}


To explicitly characterize the dependence of the equivalent channel
on the ME-STARS coefficients, we introduce
$\mathbf D_u(\mathbf Q)\triangleq
\operatorname{diag}\bigl(\mathbf h_{{\rm s},u}(\mathbf Q)\bigr)$.
Using
$\boldsymbol{\Xi}_u\mathbf h_{{\rm s},u}(\mathbf Q)
=\mathbf D_u(\mathbf Q)\boldsymbol{\theta}_{x_u}$,
the estimated equivalent channel can be expressed as
\begin{equation}
	\widehat{\mathbf h}_u
	=
	\mathbf H_{\rm BS}^H(\mathbf Q)
	\mathbf D_u(\mathbf Q)
	\boldsymbol{\theta}_{x_u}.
	\label{eq:passive_equivalent_channel}
\end{equation}
For any Hermitian positive semidefinite matrix
$\mathbf S\in\mathbb C^{N_{\rm B}\times N_{\rm B}}$, we further
introduce
\begin{equation}
	\mathbf F_{u,\mathbf S}
	\triangleq
	\mathbf D_u^H(\mathbf Q)
	\mathbf H_{\rm BS}(\mathbf Q)
	\mathbf S
	\mathbf H_{\rm BS}^H(\mathbf Q)
	\mathbf D_u(\mathbf Q),
	\label{eq:passive_F_matrix}
\end{equation}
such that
\begin{equation}
	\widehat{\mathbf h}_u^H
	\mathbf S
	\widehat{\mathbf h}_u
	=
	\boldsymbol{\theta}_{x_u}^H
	\mathbf F_{u,\mathbf S}
	\boldsymbol{\theta}_{x_u}.
	\label{eq:passive_quadratic_form}
\end{equation}
The passive beamforming matrices are defined as
$\mathbf E_{\rm r}\triangleq
\boldsymbol{\theta}_{\rm r}\boldsymbol{\theta}_{\rm r}^H$
and
$\mathbf E_{\rm t}\triangleq
\boldsymbol{\theta}_{\rm t}\boldsymbol{\theta}_{\rm t}^H$.
Consequently, \eqref{eq:passive_quadratic_form} can be equivalently
written as
\begin{equation}
	\widehat{\mathbf h}_u^H
	\mathbf S
	\widehat{\mathbf h}_u
	=
	\operatorname{Tr}
	\left(
	\mathbf F_{u,\mathbf S}\mathbf E_{x_u}
	\right).
	\label{eq:passive_trace_form}
\end{equation}
For notational compactness, we denote $\mathbf F_u^{0}\triangleq \mathbf F_{u,\mathbf V_0+\mathbf V_{\rm p}}$,
$\mathbf F_u^{\rm p}\triangleq \mathbf F_{u,\mathbf V_{\rm p}}$, and $\mathbf F_u^{-}\triangleq \mathbf F_{u,\mathbf V_{-u}}$. Then, the robust common- and private-stream power terms can then
be written as
\begin{align}
	A_u^0(\mathbf E_{x_u})
	&=
	\operatorname{Tr}
	\left(
	\mathbf F_u^0
	\mathbf E_{x_u}
	\right)
	+
	\eta_u
	\operatorname{Tr}
	\left(
	\mathbf V_{\rm p}-\mathbf V_0
	\right)
	+
	\sigma_u^2,
	\label{eq:passive_A0}\\
	B_u^0(\mathbf E_{x_u})
	&=
	\operatorname{Tr}
	\left(
	\mathbf F_u^{\rm p}
	\mathbf E_{x_u}
	\right)
	+
	\eta_u
	\operatorname{Tr}
	\left(
	\mathbf V_{\rm p}
	\right)
	+
	\sigma_u^2,
	\label{eq:passive_B0}\\
	A_u^{\rm p}(\mathbf E_{x_u})
	&=
	\operatorname{Tr}
	\left(
	\mathbf F_u^{\rm p}
	\mathbf E_{x_u}
	\right)
	+
	\eta_u
	\operatorname{Tr}
	\left(
	\mathbf V_{-u}-\mathbf V_u
	\right)
	+
	\sigma_u^2,
	\label{eq:passive_Ap}\\
	B_u^{\rm p}(\mathbf E_{x_u})
	&=
	\operatorname{Tr}
	\left(
	\mathbf F_u^{-}
	\mathbf E_{x_u}
	\right)
	+
	\eta_u
	\operatorname{Tr}
	\left(
	\mathbf V_{-u}
	\right)
	+
	\sigma_u^2.
	\label{eq:passive_Bp}
\end{align}
Hence, the robust rates are
\begin{align}
	\overline R_u^0(\mathbf E_{x_u})
	&=
	\log_2
	A_u^0(\mathbf E_{x_u})
	-
	\log_2
	B_u^0(\mathbf E_{x_u}),
	\label{eq:passive_common_rate}\\
	\overline R_u^{\rm p}(\mathbf E_{x_u})
	&=
	\log_2
	A_u^{\rm p}(\mathbf E_{x_u})
	-
	\log_2
	B_u^{\rm p}(\mathbf E_{x_u}).
	\label{eq:passive_private_rate}
\end{align}

Accordingly, the ME-STARS passive beamforming subproblem can be reformulated as
\begin{subequations}
	\label{eq:exact_passive_problem}
	\begin{align}
		\max_{
			\mathbf E_{\rm r},
			\mathbf E_{\rm t},
			\mathbf c
		}\quad
		&
		R_0(\mathbf c)
		+
		\sum_{u\in\mathcal U}
		\overline R_u^{\rm p}
		(\mathbf E_{x_u})
		\label{eq:exact_passive_objective}\\
		\mathrm{s.t.}\quad
		&
		R_0(\mathbf c)
		\leq
		\overline R_u^0
		(\mathbf E_{x_u}),
		\qquad
		\forall u\in\mathcal U,
		\label{eq:exact_passive_common}\\
		&
		c_u+
		\overline R_u^{\rm p}
		(\mathbf E_{x_u})
		\geq
		R_u^{\min},
		\qquad
		\forall u\in\mathcal U,
		\label{eq:exact_passive_QoS}\\
		&
		\operatorname{diag}(\mathbf E_{\rm r})
		+
		\operatorname{diag}(\mathbf E_{\rm t})
		\leq
		\mathbf 1_{N_{\rm S}},
		\label{eq:exact_passive_energy}\\
		&
		\mathbf E_{\rm r}\succeq\mathbf 0,
		\qquad
		\mathbf E_{\rm t}\succeq\mathbf 0,
		\label{eq:exact_passive_PSD}\\
		&
		\operatorname{rank}(\mathbf E_{\rm r})
		\leq1,
		\qquad
		\operatorname{rank}(\mathbf E_{\rm t})
		\leq1,
		\label{eq:exact_passive_rank}\\
		&
		c_u\geq0,
		\qquad
		\forall u\in\mathcal U.
		\label{eq:exact_passive_common_nonnegative}
	\end{align}
\end{subequations}

Problem \eqref{eq:exact_passive_problem} is non-convex due to the difference-of-logarithm rate functions and the rank-one constraints. To track its convexity, at iteration $\ell$, let $\mathbf E_{\rm r}^{(\ell)}$ and $\mathbf E_{\rm t}^{(\ell)}$ denote the current feasible points, and define $B_u^{0,(\ell)}\triangleq B_u^0 (\mathbf E_{x_u}^{(\ell)})$, $B_u^{{\rm p},(\ell)} \triangleq B_u^{\rm p}(\mathbf E_{x_u}^{(\ell)})$.

Since $\log_2(x)$ is concave, its first-order Taylor
expansion gives a global affine upper bound. Therefore,
\begin{align}
	\log_2
	B_u^0(\mathbf E_{x_u})
	&\leq
	\Omega_u^{0,(\ell)}
	(\mathbf E_{x_u}),
	\label{eq:passive_log_B0_upper}\\
	\log_2
	B_u^{\rm p}(\mathbf E_{x_u})
	&\leq
	\Omega_u^{{\rm p},(\ell)}
	(\mathbf E_{x_u}),
	\label{eq:passive_log_Bp_upper}
\end{align}
where
\begin{align}
	\Omega_u^{0,(\ell)}
	(\mathbf E_{x_u})
	&\triangleq
	\log_2 B_u^{0,(\ell)}
	+
	\frac{
		\operatorname{Tr}
		\left[
		\mathbf F_u^{\rm p}
		\left(
		\mathbf E_{x_u}
		-
		\mathbf E_{x_u}^{(\ell)}
		\right)
		\right]
	}{
		B_u^{0,(\ell)}\ln2
	},
	\label{eq:passive_Omega0}\\
	\Omega_u^{{\rm p},(\ell)}
	(\mathbf E_{x_u})
	&\triangleq
	\log_2 B_u^{{\rm p},(\ell)}
	+
	\frac{
		\operatorname{Tr}
		\left[
		\mathbf F_u^{-}
		\left(
		\mathbf E_{x_u}
		-
		\mathbf E_{x_u}^{(\ell)}
		\right)
		\right]
	}{
		B_u^{{\rm p},(\ell)}\ln2
	}.
	\label{eq:passive_Omegap}
\end{align}

Consequently, concave lower bounds of the robust common-
and private-stream rates are
\begin{align}
	\widetilde R_u^{0,(\ell)}
	(\mathbf E_{x_u})
	&\triangleq
	\log_2
	A_u^0(\mathbf E_{x_u})
	-
	\Omega_u^{0,(\ell)}
	(\mathbf E_{x_u}),
	\label{eq:passive_common_rate_lower}\\
	\widetilde R_u^{{\rm p},(\ell)}
	(\mathbf E_{x_u})
	&\triangleq
	\log_2
	A_u^{\rm p}(\mathbf E_{x_u})
	-
	\Omega_u^{{\rm p},(\ell)}
	(\mathbf E_{x_u}).
	\label{eq:passive_private_rate_lower}
\end{align}
The resulting approximations satisfy $\widetilde R_u^{0,(\ell)}(\mathbf E_{x_u}) \leq \overline R_u^0(\mathbf E_{x_u})$ and $\widetilde R_u^{{\rm p},(\ell)}(\mathbf E_{x_u}) \leq \overline R_u^{\rm p}(\mathbf E_{x_u})$,
$\forall u\in\mathcal U$. Moreover, both approximations are value- and gradient-consistent with their respective original rate functions at the current iterate $\mathbf E_{x_u}=\mathbf E_{x_u}^{(\ell)}$.

To facilitate the passive beamforming reformulation, we further introduce an auxiliary variable $\zeta_u$ to represent the total rate of user $u$. After relaxing the rank-one constraints in \eqref{eq:exact_passive_rank}, the resulting ME-STARS passive beamforming subproblem is formulated as
\begin{subequations}
	\label{eq:convex_passive_problem}
	\begin{align}
		\max_{\mathcal Y_{\rm P}}\quad
		&
		\sum_{u\in\mathcal U}\zeta_u
		\label{eq:convex_passive_objective}\\
		\mathrm{s.t.}\quad
		&
		\zeta_u
		\leq
		c_u+
		\widetilde R_u^{{\rm p},(\ell)}
		(\mathbf E_{x_u}),
		\qquad
		\forall u\in\mathcal U,
		\label{eq:convex_passive_total_rate}\\
		&
		\zeta_u
		\geq
		R_u^{\min},
		\qquad
		\forall u\in\mathcal U,
		\label{eq:convex_passive_QoS}\\
		&
		R_0(\mathbf c)
		\leq
		\widetilde R_u^{0,(\ell)}
		(\mathbf E_{x_u}),
		\qquad
		\forall u\in\mathcal U,
		\label{eq:convex_passive_common}\\
		&
		A_u^0(\mathbf E_{x_u})
		\geq
		\xi,
		\qquad
		\forall u\in\mathcal U,
		\label{eq:convex_passive_A0_positive}\\
		&
		A_u^{\rm p}(\mathbf E_{x_u})
		\geq
		\xi,
		\qquad
		\forall u\in\mathcal U,
		\label{eq:convex_passive_Ap_positive}\\
		&
		\operatorname{diag}(\mathbf E_{\rm r})
		+
		\operatorname{diag}(\mathbf E_{\rm t})
		\leq
		\mathbf 1_{N_{\rm S}},
		\label{eq:convex_passive_energy}
	    \end{align}
		\begin{align}
		&\mathbf E_{\rm r}\succeq\mathbf 0,
		\qquad
		\mathbf E_{\rm t}\succeq\mathbf 0,
		\label{eq:convex_passive_PSD}\\
		&
		c_u\geq0,
		\qquad
		\forall u\in\mathcal U,
		\label{eq:convex_passive_common_nonnegative}
	\end{align}
\end{subequations}
where $\xi>0$ is a sufficiently small constant and $\mathcal Y_{\rm P}\triangleq\left\{\mathbf E_{\rm r}, \mathbf E_{\rm t},\mathbf c,\{\zeta_u\}_{u\in\mathcal U}\right\}$. Problem \eqref{eq:convex_passive_problem} is convex, since the objective function is affine, the constraints in \eqref{eq:convex_passive_total_rate} and
\eqref{eq:convex_passive_common} involve concave rate approximations, and the remaining constraints are either affine or semidefinite. Therefore, the resulting subproblem can be efficiently solved using standard convex optimization solvers.

Let $\mathbf E_{\rm r}^{\star}$, $\mathbf E_{\rm t}^{\star}$, and $\mathbf c^\star$ denote the
optimal solution of \eqref{eq:convex_passive_problem}. The corresponding variables for the next iteration are updated as
\begin{equation}
	\mathbf E_{\rm r}^{(\ell+1)}
	=
	\mathbf E_{\rm r}^{\star},
	\qquad
	\mathbf E_{\rm t}^{(\ell+1)}
	=
	\mathbf E_{\rm t}^{\star},
	\qquad
	\mathbf c^{(\ell+1)}
	=
	\mathbf c^\star.
	\label{eq:passive_SCA_update}
\end{equation}
The procedure is repeated until convergence. If the optimized matrices are rank one, the reflection and
transmission coefficient vectors are recovered as
\begin{align}
	\boldsymbol{\theta}_{\rm r}^{\star}
	&=
	\sqrt{
		\lambda_{\max}
		\left(
		\mathbf E_{\rm r}^{\star}
		\right)}
	\,
	\mathbf u_{\max}
	\left(
	\mathbf E_{\rm r}^{\star}
	\right),
	\nonumber\\
	\boldsymbol{\theta}_{\rm t}^{\star}
	&=
	\sqrt{
		\lambda_{\max}
		\left(
		\mathbf E_{\rm t}^{\star}
		\right)}
	\,
	\mathbf u_{\max}
	\left(
	\mathbf E_{\rm t}^{\star}
	\right),
	\label{eq:passive_rank_one_recovery}
\end{align}
where $\lambda_{\max}(\cdot)$ and $\mathbf u_{\max}(\cdot)$ denote the largest eigenvalue and
its corresponding unit-norm eigenvector, respectively.

The amplitude and phase coefficients are then obtained as
\begin{align}
	\rho_n^\star
	&=
	\left|
	[\boldsymbol{\theta}_{\rm r}^{\star}]_n
	\right|,
	&
	\omega_n^\star
	&=
	\operatorname{mod}
	\left(
	\arg
	([\boldsymbol{\theta}_{\rm r}^{\star}]_n),
	2\pi
	\right),
	\nonumber\\
	\tau_n^\star
	&=
	\left|
	[\boldsymbol{\theta}_{\rm t}^{\star}]_n
	\right|,
	&
	\varphi_n^\star
	&=
	\operatorname{mod}
	\left(
	\arg
	([\boldsymbol{\theta}_{\rm t}^{\star}]_n),
	2\pi
	\right).
	\label{eq:passive_coefficient_recovery}
\end{align}

If either $\mathbf E_{\rm r}^{\star}$ or $\mathbf E_{\rm t}^{\star}$ has rank greater than one, Gaussian randomization is employed to construct candidate reflection and transmission vectors \cite{ni2021resource}. For each randomized candidate pair $(\widetilde{\boldsymbol{\theta}}_{\rm r}, \widetilde{\boldsymbol{\theta}}_{\rm t})$, the element-wise normalization
\begin{equation}
	s_n
	\triangleq
	\max
	\left\{
	1,
	\sqrt{
		\left|
		[\widetilde{\boldsymbol{\theta}}_{\rm r}]_n
		\right|^2
		+
		\left|
		[\widetilde{\boldsymbol{\theta}}_{\rm t}]_n
		\right|^2
	}
	\right\}
	\label{eq:passive_randomization_scaling}
\end{equation}
is applied as
\begin{equation}
	[\boldsymbol{\theta}_{\rm r}]_n
	=
	\frac{
		[\widetilde{\boldsymbol{\theta}}_{\rm r}]_n
	}{s_n},
	\qquad
	[\boldsymbol{\theta}_{\rm t}]_n
	=
	\frac{
		[\widetilde{\boldsymbol{\theta}}_{\rm t}]_n
	}{s_n},
	\label{eq:passive_randomization_normalization}
\end{equation}
thereby guaranteeing $\rho_n^2+\tau_n^2\leq1$. Among all feasible randomized candidates, the one yielding the largest original robust sum rate is selected.

\subsection{Optimization of ME-STARS Element Positions}
For the ME-STARS element-position optimization, the BS
beamforming vectors and the reflection/transmission coefficients
are kept fixed, while the element-position matrix $\mathbf Q$
and the common-rate allocation vector $\mathbf c$ are jointly
optimized. To explicitly characterize the dependence of the
equivalent cascaded channel on $\mathbf Q$, we define
\begin{equation}
	\widehat{\mathbf h}_u(\mathbf Q)
	\triangleq
	\widehat{\mathbf g}_u^H(\mathbf Q)
	=
	\mathbf H_{\rm BS}^H(\mathbf Q)
	\boldsymbol{\Xi}_u
	\mathbf h_{{\rm s},u}(\mathbf Q).
	\label{eq:column_equivalent_channel}
\end{equation}
Accordingly, the corresponding received-power terms are expressed as
\begin{align}
	A_u^0(\mathbf Q)
	&=
	\widehat{\mathbf h}_u^H(\mathbf Q)
	\left(\mathbf V_0+\mathbf V_{\rm p}\right)
	\widehat{\mathbf h}_u(\mathbf Q)
	+
	\eta_u\operatorname{Tr}
	\left(\mathbf V_{\rm p}-\mathbf V_0\right)
	+
	\sigma_u^2,
	\label{eq:position_A0}\\
	B_u^0(\mathbf Q)
	&=
	\widehat{\mathbf h}_u^H(\mathbf Q)
	\mathbf V_{\rm p}
	\widehat{\mathbf h}_u(\mathbf Q)
	+
	\eta_u\operatorname{Tr}(\mathbf V_{\rm p})
	+
	\sigma_u^2,
	\label{eq:position_B0},
		\end{align}
	\begin{align}
	A_u^{\rm p}(\mathbf Q)
	&\scalemath{0.95}{=
	\widehat{\mathbf h}_u^H(\mathbf Q)
	\mathbf V_{\rm p}
	\widehat{\mathbf h}_u(\mathbf Q)
	+
	\eta_u\operatorname{Tr}
	\left(\mathbf V_{-u}-\mathbf V_u\right)
	+
	\sigma_u^2,}
	\label{eq:position_Ap}\\
	B_u^{\rm p}(\mathbf Q)
	&=
	\widehat{\mathbf h}_u^H(\mathbf Q)
	\mathbf V_{-u}
	\widehat{\mathbf h}_u(\mathbf Q)
	+
	\eta_u\operatorname{Tr}(\mathbf V_{-u})
	+
	\sigma_u^2.
	\label{eq:position_Bp}
\end{align}

The corresponding robust common- and private-stream rates are then given by
\begin{align}
	\overline R_u^0(\mathbf Q)
	&=
	\log_2 A_u^0(\mathbf Q)
	-
	\log_2 B_u^0(\mathbf Q),
	\nonumber\\
	\overline R_u^{\rm p}(\mathbf Q)
	&=
	\log_2 A_u^{\rm p}(\mathbf Q)
	-
	\log_2 B_u^{\rm p}(\mathbf Q).
	\label{eq:position_log_difference}
\end{align}
Accordingly, the ME-STARS position-update subproblem formulated as
\begin{subequations}
	\label{eq:position_problem}
	\begin{align}
		\max_{\mathbf Q,\mathbf c}\quad
		&
		R_0(\mathbf c)
		+
		\sum_{u\in\mathcal U}
		\overline R_u^{\rm p}(\mathbf Q)
		\label{eq:position_objective}\\
		\mathrm{s.t.}\quad
		&
		R_0(\mathbf c)
		\leq
		\overline R_u^0(\mathbf Q),
		\qquad \forall u\in\mathcal U,
		\label{eq:position_common}\\
		&
		c_u+\overline R_u^{\rm p}(\mathbf Q)
		\geq
		R_u^{\min},
		\qquad \forall u\in\mathcal U,
		\label{eq:position_QoS}\\
		&
		c_u\geq0,
		\qquad \forall u\in\mathcal U,
		\label{eq:position_common_nonnegative}\\
		&
		\mathbf q_n\in\mathcal C,
		\qquad \forall n,
		\label{eq:position_region}\\
		&
		\|\mathbf q_n-\mathbf q_m\|_2\geq d_{\min},
		\qquad \forall n\neq m.
		\label{eq:position_spacing}
	\end{align}
\end{subequations}
Problem \eqref{eq:position_problem} is non-convex due to the nonlinear dependence of the achievable rates on the ME-STARS element positions $\mathbf Q$ and the non-convex minimum inter-element spacing constraint in \eqref{eq:position_spacing}.

\subsubsection{Sequential Single-Element Update}
The ME-STARS elements are updated sequentially. When optimizing the $n$-th element, the positions of all remaining elements $\{\mathbf q_m\}_{m\neq n}$ are kept fixed, while $\mathbf q_n$ and the common-rate allocation vector $\mathbf c$ are jointly optimized. Let $\mathbf q_n^{(\ell)}$ and $\mathbf c^{(\ell)}$ denote the current feasible solutions at inner iteration $\ell$, and define the position displacement as
$\Delta\mathbf q_n\triangleq \mathbf q_n-\mathbf q_n^{(\ell)}$.

For notational convenience, we introduce the direction-dependent
vectors
\begin{equation}
	\mathbf d(\vartheta,\varpi)
	\triangleq
	\begin{bmatrix}
		\sin(\vartheta)\cos(\varpi)\\
		\sin(\varpi)
	\end{bmatrix},
	\qquad
	\boldsymbol{\nu}(\vartheta,\varpi)
	\triangleq
	k_0\mathbf d(\vartheta,\varpi).
	\label{eq:direction_vector}
\end{equation}
Accordingly, let $\boldsymbol{\nu}_{\rm in}\triangleq \boldsymbol{\nu}(\vartheta_{\rm in},\varpi_{\rm in})$
and $\boldsymbol{\nu}_u\triangleq \boldsymbol{\nu}(\vartheta_u,\varpi_u)$, where $(\vartheta_u,\varpi_u)$ denote the azimuth and elevation angles associated with the ME-STARS--user $u$ link. The position-dependent coefficients corresponding to the $n$-th movable element can then be expressed as
\begin{equation}
	u_{{\rm in},n}
	=
	e^{j\boldsymbol{\nu}_{\rm in}^T\mathbf q_n},
	\qquad
	h_{{\rm s},u,n}
	=
	\sqrt{\ell_u}\,
	e^{j\boldsymbol{\nu}_u^T\mathbf q_n},
	\label{eq:element_coefficients}
\end{equation}
where $\ell_u$ denotes the large-scale channel gain of the corresponding ME-STARS--user link. 

Further, to simplify the subsequent derivations, let $\alpha_{\rm BS}\triangleq \sqrt{\ell_{\rm BS}\chi_{\rm BS}/(\chi_{\rm BS}+1)}$, and let $\mathbf e_n$ denote the $n$-th canonical basis vector of $\mathbb R^{N_{\rm S}}$. Since $\mathbf Z_{\rm BS}$ is independent of the ME-STARS element positions, its derivatives
with respect to $\mathbf q_n$ are zero. Hence, the first- and second-order partial derivatives of $\mathbf H_{\rm BS}(\mathbf Q)$ in \eqref{eq:H_BS} with respect to the coordinates of the $n$-th movable element are given by

\begin{align}
	\mathbf H_a
	\triangleq
	\frac{\partial\mathbf H_{\rm BS}}{\partial q_{a,n}}
	&=
	j\alpha_{\rm BS}
	\nu_{{\rm in},a}
	u_{{\rm in},n}
	\mathbf e_n\mathbf b^H(\vartheta_{\rm out}),
	\label{eq:H_first_derivative}\\
	\mathbf H_{ab}
	\triangleq
	\frac{\partial^2\mathbf H_{\rm BS}}
	{\partial q_{a,n}\partial q_{b,n}}
	&=
	-\alpha_{\rm BS}
	\nu_{{\rm in},a}\nu_{{\rm in},b}
	u_{{\rm in},n}
	\mathbf e_n\mathbf b^H(\vartheta_{\rm out}),
	\label{eq:H_second_derivative}
\end{align}
where $a,b\in\{x,y\}$. Likewise,
\begin{align}
	\mathbf h_{{\rm s},u,a}
	\triangleq
	\frac{\partial\mathbf h_{{\rm s},u}}{\partial q_{a,n}}
	&=
	j\nu_{u,a}
	h_{{\rm s},u,n}\mathbf e_n,
	\label{eq:hs_first_derivative}\\
	\mathbf h_{{\rm s},u,ab}
	\triangleq
	\frac{\partial^2\mathbf h_{{\rm s},u}}
	{\partial q_{a,n}\partial q_{b,n}}
	&=
	-\nu_{u,a}\nu_{u,b}
	h_{{\rm s},u,n}\mathbf e_n.
	\label{eq:hs_second_derivative}
\end{align}

Let
\begin{equation}
	\mathbf z_u
	\triangleq
	\boldsymbol{\Xi}_u\mathbf h_{{\rm s},u},
	\qquad
	\widehat{\mathbf h}_u
	=
	\mathbf H_{\rm BS}^H\mathbf z_u.
	\label{eq:weighted_user_channel}
\end{equation}
Applying the product rule to \eqref{eq:weighted_user_channel} gives
\begin{align}
	\widehat{\mathbf h}_{u,a}
	\triangleq
	\frac{\partial\widehat{\mathbf h}_u}{\partial q_{a,n}}
	&=
	\mathbf H_a^H\mathbf z_u
	+
	\mathbf H_{\rm BS}^H
	\boldsymbol{\Xi}_u
	\mathbf h_{{\rm s},u,a},
	\label{eq:equivalent_first_derivative}\\
	\widehat{\mathbf h}_{u,ab}
	\triangleq
	\frac{\partial^2\widehat{\mathbf h}_u}
	{\partial q_{a,n}\partial q_{b,n}}
	&=
	\mathbf H_{ab}^H\mathbf z_u
	+
	\mathbf H_a^H\boldsymbol{\Xi}_u\mathbf h_{{\rm s},u,b}+
	\mathbf H_b^H\boldsymbol{\Xi}_u\mathbf h_{{\rm s},u,a}	\nonumber\\
	&\quad
	+
	\mathbf H_{\rm BS}^H
	\boldsymbol{\Xi}_u
	\mathbf h_{{\rm s},u,ab}.
	\label{eq:equivalent_second_derivative}
\end{align}

\subsubsection{Gradient and Hessian of the Position-Dependent Terms}

The position-dependent quadratic terms in
\eqref{eq:position_A0}--\eqref{eq:position_Bp} have the generic quadratic form
\begin{equation}
	f_{u,\mathbf S}(\mathbf q_n)
	\triangleq
	\widehat{\mathbf h}_{u}^{H}(\mathbf q_n)
	\mathbf S
	\widehat{\mathbf h}_{u}(\mathbf q_n),
	\label{eq:generic_power}
\end{equation}
where $\mathbf S\succeq\mathbf 0$ is a Hermitian weighting matrix. For the
common- and private-stream power terms, the corresponding matrices are
\begin{align}
	\mathbf S_{A,u}^{0}
	&=
	\mathbf V_0+\mathbf V_{\rm p},
	&
	\mathbf S_{B,u}^{0}
	&=
	\mathbf V_{\rm p},
	\label{eq:power_weighting_common}\\
	\mathbf S_{A,u}^{\rm p}
	&=
	\mathbf V_{\rm p},
	&
	\mathbf S_{B,u}^{\rm p}
	&=
	\mathbf V_{-u}.
	\label{eq:power_weighting_private}
\end{align}
For $a,b\in\{x,y\}$, the first-order partial derivative of
\eqref{eq:generic_power} with respect to $q_{a,n}$ is
\begin{equation}
	\frac{\partial f_{u,\mathbf S}}
	{\partial q_{a,n}}
	=
	2\Re
	\left\{
	\widehat{\mathbf h}_{u,a}^{H}
	\mathbf S
	\widehat{\mathbf h}_{u}
	\right\}.
	\label{eq:power_gradient_component}
\end{equation}
whereas its second-order partial derivative is
\begin{equation}
	\frac{\partial^2 f_{u,\mathbf S}}
	{\partial q_{a,n}\partial q_{b,n}}
	=
	2\Re
	\left\{
	\widehat{\mathbf h}_{u,a}^{H}
	\mathbf S
	\widehat{\mathbf h}_{u,b}
	+
	\widehat{\mathbf h}_{u}^{H}
	\mathbf S
	\widehat{\mathbf h}_{u,ab}
	\right\}.
	\label{eq:power_Hessian_component}
\end{equation}
Accordingly, the gradient vector of $f_{u,\mathbf S}$ with respect to the
position of the $n$-th element is
\begin{equation}
	\nabla f_{u,\mathbf S}(\mathbf q_n)
	=
	\begin{bmatrix}
		\dfrac{\partial f_{u,\mathbf S}}{\partial q_{x,n}}\\[2mm]
		\dfrac{\partial f_{u,\mathbf S}}{\partial q_{y,n}}
	\end{bmatrix},
	\label{eq:power_gradient_vector}
\end{equation}
and its $2\times2$ Hessian matrix is
\begin{equation}
	\nabla^2 f_{u,\mathbf S}(\mathbf q_n)
	=
	\begin{bmatrix}
		\dfrac{\partial^2 f_{u,\mathbf S}}{\partial q_{x,n}^{2}}
		&
		\dfrac{\partial^2 f_{u,\mathbf S}}
		{\partial q_{x,n}\partial q_{y,n}}
		\\[3mm]
		\dfrac{\partial^2 f_{u,\mathbf S}}
		{\partial q_{y,n}\partial q_{x,n}}
		&
		\dfrac{\partial^2 f_{u,\mathbf S}}{\partial q_{y,n}^{2}}
	\end{bmatrix}.
	\label{eq:power_Hessian_matrix}
\end{equation}
Hence, the gradients and Hessians of
$A_u^{0}(\mathbf q_n)$, $B_u^{0}(\mathbf q_n)$,
$A_u^{\rm p}(\mathbf q_n)$, and $B_u^{\rm p}(\mathbf q_n)$
are obtained directly from
\eqref{eq:power_gradient_vector} and
\eqref{eq:power_Hessian_matrix} by substituting the
corresponding weighting matrices defined in
\eqref{eq:power_weighting_common} and
\eqref{eq:power_weighting_private}.

\subsubsection{MM Lower and Upper Surrogates}
The position-dependent power functions $A_u^0(\mathbf q_n)$, $B_u^0(\mathbf q_n)$, $A_u^{\rm p}(\mathbf q_n)$, and $B_u^{\rm p}(\mathbf q_n)$ are twice continuously differentiable but are generally neither convex nor concave with respect to $\mathbf q_n$. To handle this non-convexity, we employ an MM framework \cite{sun2016majorization} and construct suitable concave lower and convex upper quadratic surrogates based on the following standard lemma.

\noindent\textbf{Lemma 1:}
Let $f:\mathcal C\rightarrow\mathbb R$ be a twice continuously differentiable function over the convex set
$\mathcal C$, and let $\mathbf q_n^{(\ell)}\in\mathcal C$ denote the current feasible point. If there exists a curvature constant $\delta_f^{(\ell)}\geq 0$ such that
\begin{equation}
	-\delta_f^{(\ell)}\mathbf I_2
	\preceq
	\nabla^2 f(\mathbf q)
	\preceq
	\delta_f^{(\ell)}\mathbf I_2,
	\qquad
	\forall\mathbf q\in\mathcal C,
	\label{eq:valid_curvature}
\end{equation}
then, for every $\mathbf q_n\in\mathcal C$, a concave quadratic lower bound and a convex quadratic upper bound for $f(\mathbf q_n)$ around $\mathbf q_n^{(\ell)}$ are, respectively, given by
\begin{align}
	\underline f^{(\ell)}(\mathbf q_n)
	&\triangleq
	f(\mathbf q_n^{(\ell)})
	+
	\nabla f(\mathbf q_n^{(\ell)})^T
	\Delta\mathbf q_n
	-
	\frac{\delta_f^{(\ell)}}{2}
	\|\Delta\mathbf q_n\|_2^2,
	\label{eq:MM_lower}\\
	\overline f^{(\ell)}(\mathbf q_n)
	&\triangleq
	f(\mathbf q_n^{(\ell)})
	+
	\nabla f(\mathbf q_n^{(\ell)})^T
	\Delta\mathbf q_n
	+
	\frac{\delta_f^{(\ell)}}{2}
	\|\Delta\mathbf q_n\|_2^2.
	\label{eq:MM_upper}
\end{align}
Moreover, the resulting surrogate functions satisfy $\underline f^{(\ell)}(\mathbf q_n) \leq f(\mathbf q_n) \leq \overline f^{(\ell)}(\mathbf q_n)$, $\forall\,\mathbf q_n\in\mathcal C$, and both surrogates are value- and gradient-consistent with the original function at $\mathbf q_n^{(\ell)}$. In practice, a curvature estimate can be initialized from
$\|\nabla^2 f(\mathbf q_n^{(\ell)})\|_{\rm F}$ and
adaptively increased until the required surrogate-bound and
acceptance conditions are satisfied.

Accordingly, applying
\eqref{eq:MM_lower}--\eqref{eq:MM_upper} to the position-dependent terms gives
\begin{align}
	\underline A_u^{0,(\ell)}(\mathbf q_n)
	&\scalemath{0.88}{=
	A_u^0(\mathbf q_n^{(\ell)})
	+
	\nabla A_u^0(\mathbf q_n^{(\ell)})^T
	\Delta\mathbf q_n
	-
	\frac{\delta_{A,u}^{0,(\ell)}}{2}
	\|\Delta\mathbf q_n\|_2^2,}
	\label{eq:A0_lower}\\
	\overline B_u^{0,(\ell)}(\mathbf q_n)
	&\scalemath{0.88}{=
	B_u^0(\mathbf q_n^{(\ell)})
	+
	\nabla B_u^0(\mathbf q_n^{(\ell)})^T
	\Delta\mathbf q_n
	+
	\frac{\delta_{B,u}^{0,(\ell)}}{2}
	\|\Delta\mathbf q_n\|_2^2,}
	\label{eq:B0_upper}\\
	\underline A_u^{{\rm p},(\ell)}(\mathbf q_n)
	&\scalemath{0.88}{=
	A_u^{\rm p}(\mathbf q_n^{(\ell)})
	+
	\nabla A_u^{\rm p}(\mathbf q_n^{(\ell)})^T
	\Delta\mathbf q_n
	-
	\frac{\delta_{A,u}^{{\rm p},(\ell)}}{2}
	\|\Delta\mathbf q_n\|_2^2,}
	\label{eq:Ap_lower}\\
	\overline B_u^{{\rm p},(\ell)}(\mathbf q_n)
	&\scalemath{0.88}{=
	B_u^{\rm p}(\mathbf q_n^{(\ell)})
	+
	\nabla B_u^{\rm p}(\mathbf q_n^{(\ell)})^T
	\Delta\mathbf q_n
	+
	\frac{\delta_{B,u}^{{\rm p},(\ell)}}{2}
	\|\Delta\mathbf q_n\|_2^2.}
	\label{eq:Bp_upper}
\end{align}
Here, the curvature parameter associated with each power function is chosen to satisfy the condition in \eqref{eq:valid_curvature}.

\subsubsection{Concave Lower Bounds of the Robust Rates}
To facilitate the rate approximation, we introduce auxiliary variables $\alpha_u^0\geq\xi$ and $\alpha_u^{\rm p}\geq\xi$, where $\xi>0$ is a sufficiently small constant, and require $\alpha_u^0\leq\underline A_u^{0,(\ell)}(\mathbf q_n)$ and $\alpha_u^{\rm p}\leq\underline A_u^{{\rm p},(\ell)}(\mathbf q_n)$,
$\forall u\in\mathcal U$. Since $\log_2(x)$ is monotonically increasing and concave, and
$B_u^0(\mathbf q_n)\leq\overline B_u^{0,(\ell)}(\mathbf q_n)$ and
$B_u^{\rm p}(\mathbf q_n)\leq\overline B_u^{{\rm p},(\ell)}(\mathbf q_n)$,
we further define
$B_u^{0,(\ell)}\triangleq B_u^0(\mathbf q_n^{(\ell)})$ and
$B_u^{{\rm p},(\ell)}\triangleq
B_u^{\rm p}(\mathbf q_n^{(\ell)})$.

Accordingly, the first-order upper bounds are 
\begin{align}
	\log_2 B_u^0(\mathbf q_n)
	&\leq
	\log_2 B_u^{0,(\ell)}
	+
	\frac{
		\overline B_u^{0,(\ell)}(\mathbf q_n)-B_u^{0,(\ell)}
	}{
		B_u^{0,(\ell)}\ln2
	},
	\label{eq:log_B0_upper}\\
	\log_2 B_u^{\rm p}(\mathbf q_n)
	&\leq
	\log_2 B_u^{{\rm p},(\ell)}
	+
	\frac{
		\overline B_u^{{\rm p},(\ell)}(\mathbf q_n)-B_u^{{\rm p},(\ell)}
	}{
		B_u^{{\rm p},(\ell)}\ln2
	}.
	\label{eq:log_Bp_upper}
\end{align}
Consequently, concave lower bounds of the common- and private-stream rates are
\begin{align}
	\widetilde R_u^{0,(\ell)}(\mathbf q_n,\alpha_u^0)
	&=
	\log_2\alpha_u^0
	-
	\log_2B_u^{0,(\ell)}
	-
	\frac{
		\overline B_u^{0,(\ell)}(\mathbf q_n)-B_u^{0,(\ell)}
	}{
		B_u^{0,(\ell)}\ln2
	},
	\label{eq:common_rate_lower}\\
	\widetilde R_u^{{\rm p},(\ell)}(\mathbf q_n,\alpha_u^{\rm p})
	&=
	\log_2\alpha_u^{\rm p}
	-
	\log_2B_u^{{\rm p},(\ell)}
	-
	\frac{
		\overline B_u^{{\rm p},(\ell)}(\mathbf q_n)-B_u^{{\rm p},(\ell)}
	}{
		B_u^{{\rm p},(\ell)}\ln2
	}.
	\label{eq:private_rate_lower}
\end{align}
These bounds satisfy
\[
\widetilde R_u^{0,(\ell)}(\mathbf q_n,\alpha_u^0)
\leq
\overline R_u^0(\mathbf q_n),
\qquad
\widetilde R_u^{{\rm p},(\ell)}(\mathbf q_n,\alpha_u^{\rm p})
\leq
\overline R_u^{\rm p}(\mathbf q_n),
\]
with equality and first-order consistency at
$\mathbf q_n=\mathbf q_n^{(\ell)}$ when the auxiliary variables are tight.

\subsubsection{Convexification of the Spacing Constraint}
For $m\neq n$, define the unit vector
\begin{equation}
	\mathbf a_{n,m}^{(\ell)}
	\triangleq
	\frac{
		\mathbf q_n^{(\ell)}-\mathbf q_m
	}{
		\left\|
		\mathbf q_n^{(\ell)}-\mathbf q_m
		\right\|_2
	}.
	\label{eq:spacing_direction}
\end{equation}
For fixed $\mathbf q_m$, the convex function
$\|\mathbf q_n-\mathbf q_m\|_2$ admits the affine lower bound
\begin{equation}
	\|\mathbf q_n-\mathbf q_m\|_2
	\geq
	\left(\mathbf a_{n,m}^{(\ell)}\right)^T
	\left(\mathbf q_n-\mathbf q_m\right).
	\label{eq:spacing_lower_bound}
\end{equation}
Therefore, the conservative affine replacement of
\eqref{eq:position_spacing} is
\begin{equation}
	\left(\mathbf a_{n,m}^{(\ell)}\right)^T
	\left(\mathbf q_n-\mathbf q_m\right)
	\geq d_{\min},
	\qquad
	\forall m\neq n.
	\label{eq:spacing_SCA}
\end{equation}

\begin{algorithm}[t]
	\caption{Successive Optimization of ME-STARS Element Positions}
	\label{alg:ME_STARS_position_optimization}
	\begin{algorithmic}[1]
		\State \textbf{Input:} Fixed $\mathcal V$,
		$\boldsymbol{\Xi}_{\rm r}$,
		$\boldsymbol{\Xi}_{\rm t}$,
		$\epsilon_{\rm Q}$, and
		${\mathcal I}_{\rm Q}^{\max}$.
		
		\State \textbf{Initialization:} Set $n=1$ and initialize
		feasible $\mathbf Q$ and $\mathbf c$.
		
		\Repeat
		
		\State Fix $\{\mathbf q_m\}_{m\neq n}$ and set $\ell=0$.
		
		\Repeat
		
		\State Compute the derivatives and curvature bounds.
		
		\State Construct the rate and spacing surrogates.
		
		\State Solve \eqref{eq:convex_position_problem} to obtain
		$\mathbf q_n^\star$ and $\mathbf c^\star$.
		
		\State Apply \eqref{eq:joint_backtracking} and update
		$\mathbf q_n^{(\ell+1)}$ and $\mathbf c^{(\ell+1)}$.
		
		\State Set $\ell\leftarrow\ell+1$.
		
		\Until{convergence or
			$\ell={\mathcal I}_{\rm Q}^{\max}$.}
		
		\State Update the $n$-th column of $\mathbf Q$.
		\State Set $n\leftarrow n+1$.
		
		\Until{$n>N_{\rm S}$.}
		
		\State \Return $\mathbf Q^\star$ and $\mathbf c^\star$.
	\end{algorithmic}
\end{algorithm}

\subsubsection{Convex Single-Element Position Subproblem}

At inner iteration $\ell$, the position of element $n$ and the common-rate allocation are jointly updated by solving
\begin{subequations}
	\label{eq:convex_position_problem}
	\begin{align}
		\max_{
			\mathbf q_n,\mathbf c,
			\{\alpha_u^0,\alpha_u^{\rm p}\}_{u\in\mathcal U}
		}\quad
		&
		R_0(\mathbf c)
		+
		\sum_{u\in\mathcal U}
		\widetilde R_u^{{\rm p},(\ell)}(\mathbf q_n,\alpha_u^{\rm p})
		\label{eq:convex_position_objective}\\
		\mathrm{s.t.}\quad
		&
		R_0(\mathbf c)
		\leq
		\widetilde R_u^{0,(\ell)}(\mathbf q_n,\alpha_u^0),
		\qquad \forall u\in\mathcal U,
		\label{eq:convex_position_common}\\
		&
		c_u+
		\widetilde R_u^{{\rm p},(\ell)}(\mathbf q_n,\alpha_u^{\rm p})
		\geq
		R_u^{\min},
		\qquad \forall u\in\mathcal U,
		\label{eq:convex_position_QoS}\\
		&
		c_u\geq0,
		\qquad \forall u\in\mathcal U,
		\label{eq:convex_position_common_nonnegative}\\
		&
		\alpha_u^0\geq\xi,\quad
		\alpha_u^{\rm p}\geq\xi,
		\qquad \forall u\in\mathcal U,
		\label{eq:convex_position_auxiliary_positive}\\
		&
		\alpha_u^0
		\leq
		\underline A_u^{0,(\ell)}(\mathbf q_n),
		\qquad \forall u\in\mathcal U,
		\label{eq:convex_position_A0}\\
		&
		\alpha_u^{\rm p}
		\leq
		\underline A_u^{{\rm p},(\ell)}(\mathbf q_n),
		\qquad \forall u\in\mathcal U,
		\label{eq:convex_position_Ap}\\
		&
		\mathbf q_n\in\mathcal C,
		\label{eq:convex_position_region}\\
		&
		\left(\mathbf a_{n,m}^{(\ell)}\right)^T
		\left(\mathbf q_n-\mathbf q_m\right)
		\geq d_{\min},
		\qquad \forall m\neq n.
		\label{eq:convex_position_spacing}
	\end{align}
\end{subequations}
Problem \eqref{eq:convex_position_problem} is convex, since its objective function is concave, the common-decoding and QoS constraints involve superlevel sets of concave functions, the auxiliary-variable constraints correspond to hypographs of concave functions, and the remaining constraints are either convex or affine.

Let $\mathbf q_n^\star$ and $\mathbf c^\star$ denote the
solution of \eqref{eq:convex_position_problem}. To safeguard
feasibility and monotonic improvement of the original robust
objective, a joint backtracking step can be employed:
\begin{align}
	\mathbf q_n(\beta)
	&=
	\mathbf q_n^{(\ell)}
	+
	\lambda
	\left(
	\mathbf q_n^\star-\mathbf q_n^{(\ell)}
	\right),
	\nonumber\\
	\mathbf c(\beta)
	&=
	\mathbf c^{(\ell)}
	+
	\lambda
	\left(
	\mathbf c^\star-\mathbf c^{(\ell)}
	\right),
	\label{eq:joint_backtracking}
\end{align}
where $\lambda\in\{1,1/2,1/4,\ldots\}$. The largest step satisfying the original robust common-decoding and QoS
constraints, while not decreasing the original robust sum rate, is accepted. After the update of element $n$ converges, the procedure proceeds to the next movable element.

\subsection{Computational Complexity Analysis}
The computational complexity of the proposed framework mainly arises from solving the active beamforming, passive beamforming, and ME-STARS element-position optimization subproblems. Let $K=|\mathcal U|=K_{\rm r}+K_{\rm t}$ denote the total number of users, while $\mathcal I_{\rm A}$, $\mathcal I_{\rm P}$, $\mathcal I_{\rm Q}$, and $\mathcal I_{\rm AO}$ denote the numbers of iterations required for the active beamforming
SCA procedure, passive beamforming SCA procedure, single-element position-optimization procedure, and overall
AO framework, respectively. The active beamforming subproblem in \eqref{eq:convex_active_problem} optimizes
$K+1$ positive semidefinite matrices of size $N_{\rm B}\times N_{\rm B}$, including one common-stream
covariance matrix and $K$ private-stream covariance matrices, its dominant computational complexity is $\mathcal O\big(\mathcal I_{\rm A}(K+1)^{3.5}N_{\rm B}^{7}\log(1/\delta_{\rm A})\big)$ \cite{luo2010semidefinite}. The passive beamforming
subproblem in \eqref{eq:convex_passive_problem} optimizes two positive semidefinite matrices of size
$N_{\rm S}\times N_{\rm S}$. Since the number of these matrices is constant, its dominant computational complexity is $\mathcal O\big(\mathcal I_{\rm P}N_{\rm S}^{7}\log(1/\delta_{\rm P})\big)$\cite{luo2010semidefinite}. Moreover, Algorithm~\ref{alg:ME_STARS_position_optimization} sequentially updates all $N_{\rm S}$ movable elements. For each selected element, the position-dependent channel derivatives and MM surrogates are constructed for all $K$
users while accounting for the remaining elements, resulting in a computational complexity of
$\mathcal O\big(\mathcal I_{\rm Q}KN_{\rm S}^{2}\log(1/\delta_{\rm Q})\big)$\cite{grant2008cvx}. Consequently, the overall computational complexity of Algorithm~\ref{alg:overall_optimization} is $\mathcal O\big(\mathcal I_{\rm AO}[
\mathcal I_{\rm A}(K+1)^{3.5}N_{\rm B}^{7}\log(1/\delta_{\rm A})+\mathcal I_{\rm P}N_{\rm S}^{7}\log(1/\delta_{\rm P})+\mathcal I_{\rm Q}KN_{\rm S}^{2}\log(1/\delta_{\rm Q})]\big)$, where$\delta_{\rm A}$, $\delta_{\rm P}$, and $\delta_{\rm Q}$ denote the prescribed numerical accuracies of the corresponding convex subproblems.

\begin{algorithm}[t]
	\caption{Overall Proposed Optimization Framework}
	\label{alg:overall_optimization}
	\begin{algorithmic}[1]
		\State \textbf{Input:} $\epsilon_{\rm AO}$ and
		${\mathcal I}_{\rm AO}^{\max}$.
		
		\State \textbf{Initialization:} Set $t=0$ and initialize
		feasible $\mathcal X^{(0)}$.
		
		\Repeat
		
		\State \textbf{Active beamforming update:}
		\State Solve \eqref{eq:convex_active_problem}.
		\State Update $\mathcal V$ and $\mathbf c$.
		\State Recover the active beamforming vectors.
		
		\State \textbf{Passive beamforming update:}
		\State Solve \eqref{eq:convex_passive_problem}.
		\State Update $\mathbf E_{\rm r}$, $\mathbf E_{\rm t}$,
		and $\mathbf c$.
		\State Recover the passive beamforming vectors.
		\State Construct $\boldsymbol{\Xi}_{\rm r}$ and
		$\boldsymbol{\Xi}_{\rm t}$.
		
		\State \textbf{Element-position update:}
		\State Apply Algorithm~\ref{alg:ME_STARS_position_optimization}.
		\State Update $\mathbf Q$, $\mathbf c$, and the channels.
		
		\State Evaluate the robust sum rate $R^{(t+1)}$.
		\State Set $t\leftarrow t+1$.
		
		\Until{convergence or
			$t={\mathcal I}_{\rm AO}^{\max}$.}
		
		\State \Return $\mathcal X^\star$.
	\end{algorithmic}
\end{algorithm}
 
\section{Numerical Results}
Unless otherwise specified, the simulations consider a BS equipped with $N_{\rm B}=8$ half-wavelength-spaced antennas, serving $K_{\rm r}=2$ reflection-region users and $K_{\rm t}=2$ transmission-region users through an ME-STARS comprising $N_{\rm S}=16$ movable elements. The carrier frequency is set to $f_{\rm c}=10$ GHz, corresponding to a wavelength of $\lambda=0.03$ m. Each movable element is confined to a square region of size
$5\lambda\times5\lambda$, and the minimum allowable distance between any two elements is set to $d_{\min}=\lambda/2$. The elements are initially placed at uniformly distributed feasible positions within the prescribed movement region. The distance between the BS and the ME-STARS is $70$ m, whereas the ME-STARS-to-transmission-user distances are $5$ m and $3$ m. The corresponding distances for the reflection-region users are $15$ m and $30$ m. The departure angle from the BS toward the ME-STARS is $\vartheta_{\rm out}=120^{\circ}$, while the azimuth and elevation angles of arrival at the ME-STARS are set to $\vartheta_{\rm in}=330^{\circ}$ and $\varpi_{\rm in}=30^{\circ}$, respectively. The transmission users are located along the directions $\boldsymbol{\vartheta}_{\rm t} =[140^{\circ},210^{\circ}]$ with elevation angles
$\boldsymbol{\varpi}_{\rm t}=[-30^{\circ},-30^{\circ}]$, whereas the reflection users are located along $\boldsymbol{\vartheta}_{\rm r}=[-45^{\circ},30^{\circ}]$ with $\boldsymbol{\varpi}_{\rm r} =[-30^{\circ},-30^{\circ}]$. The BS--ME-STARS link follows the Rician channel model with Rician factor $\chi_{\rm BS}=3$, while the ME-STARS--user links are assumed to be dominated by their LoS components \cite{liu2026joint}. The large-scale channel gain of a link with distance $d$ is modeled as $\ell(d)=\rho_0 d^{-2}$, where $\rho_0=-30$ dB denotes the reference channel gain at a distance of $1$ m. For each channel realization, the NLoS component of the BS--ME-STARS channel is generated according to $\mathcal{CN}(0,1)$. To characterize imperfect CSI, we adopt a norm-bounded CSI uncertainty model in which the uncertainty radius is scaled with its estimated channel as $\eta_u = \delta \,\big\|\widehat{\mathbf G}_u\big\|$ \cite{asif2026robust,zheng2023zero}, where $\delta \in [0,1)$ denotes the normalized CSI-uncertainty level. The receiver noise power is set to $-80$ dBm, the maximum BS transmit power is $P_{\max}=35$ dBm, and the minimum required rate of each user is $R_u^{\min}=1$ bps/Hz \cite{asif2026robust}.

\begin{figure}[!t]
	\centering
	\includegraphics [width=0.38\textwidth]{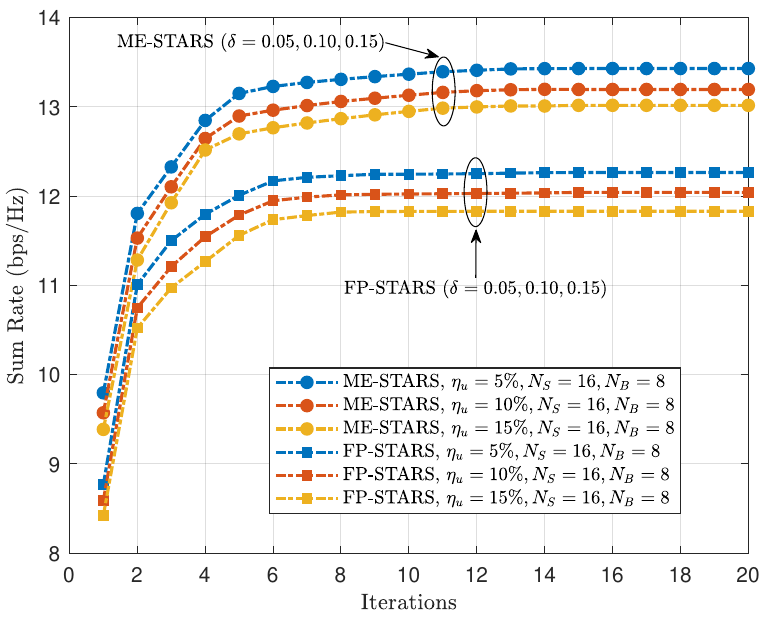}
	\caption{Convergence under different CSI uncertainty levels $\eta_u$.}
	\label{f2}
\end{figure} 

\begin{figure}[t]
	\centering
	\includegraphics [width=0.38\textwidth]{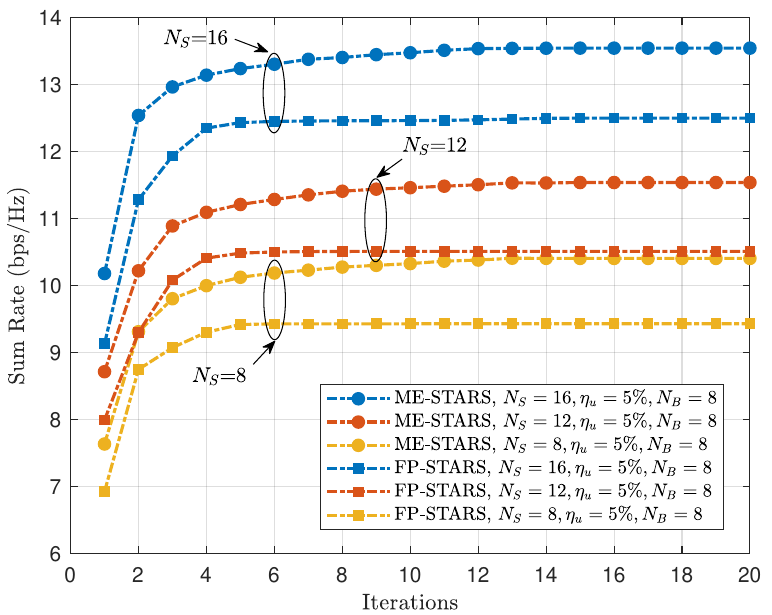}
	\caption{Convergence for different ME-STARS elements $N_{\rm S}$.}
	\label{f3}
\end{figure}

  \begin{figure}[!h]
	\centering
	\includegraphics [width=0.38\textwidth]{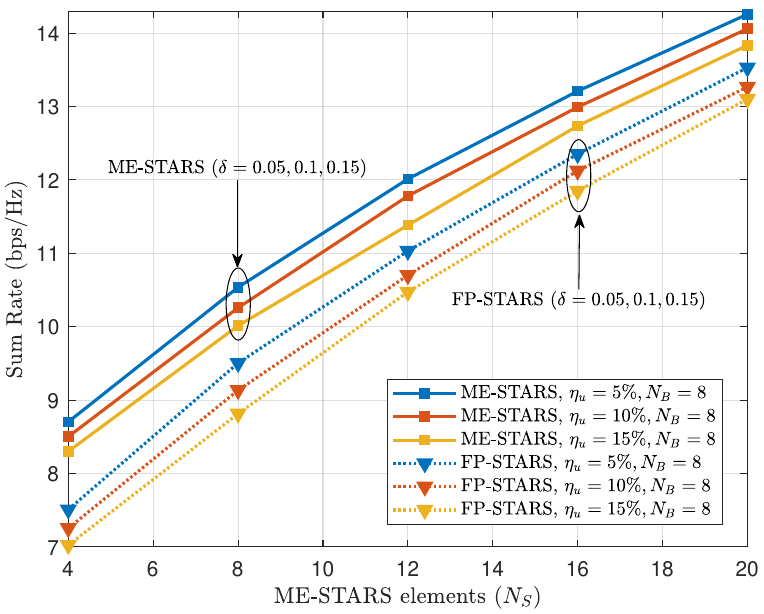}
	\caption{Sum rate versus $N_{\rm S}$ under different $\eta_u$.}
	\label{f4}
\end{figure}

\begin{figure}[h]
	\centering
	\includegraphics [width=0.38\textwidth]{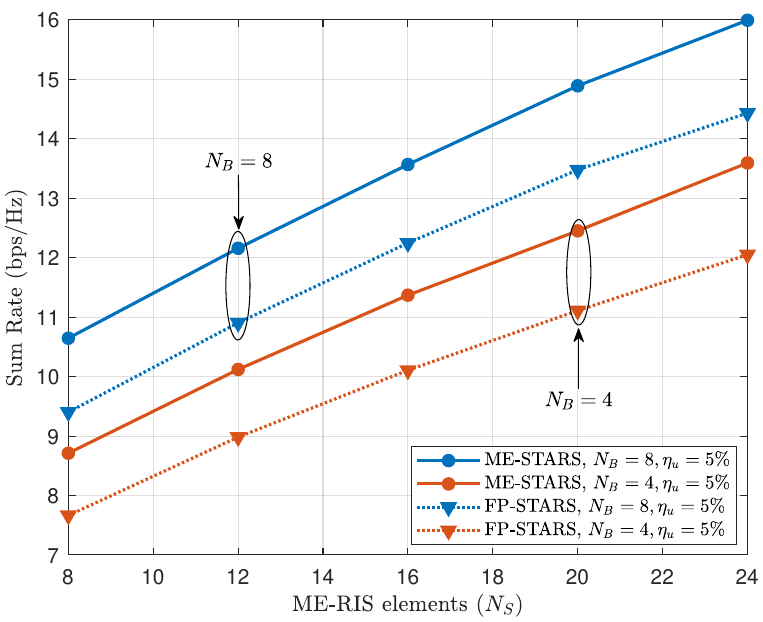}
	\caption{Sum rate versus $N_{\rm S}$ for different $N_{\rm B}$.}
	\label{f5}
\end{figure}

\begin{figure}[h]
\centering
\includegraphics [width=0.38\textwidth]{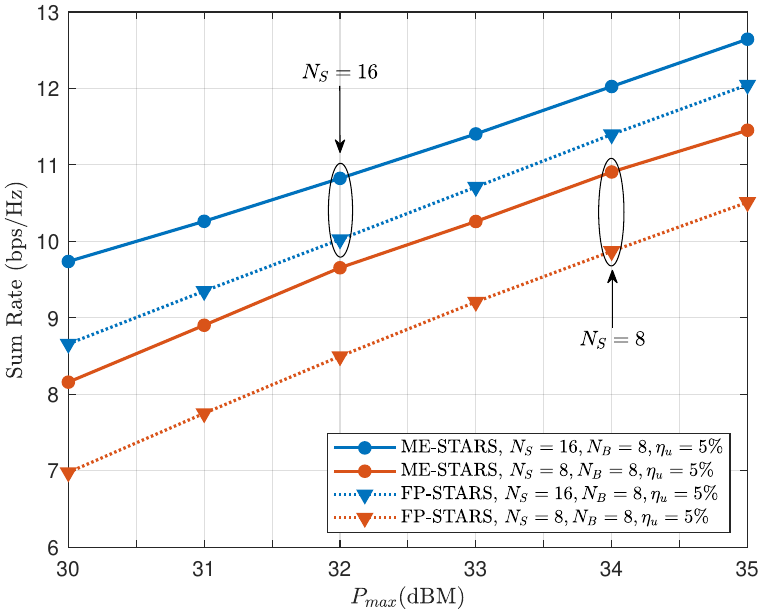}
\caption{Sum rate versus $P_{\max}$ for different $N_{\rm S}$.}
\label{f6}
\end{figure}

\begin{figure}[h]
	\centering
	\includegraphics [width=0.38\textwidth]{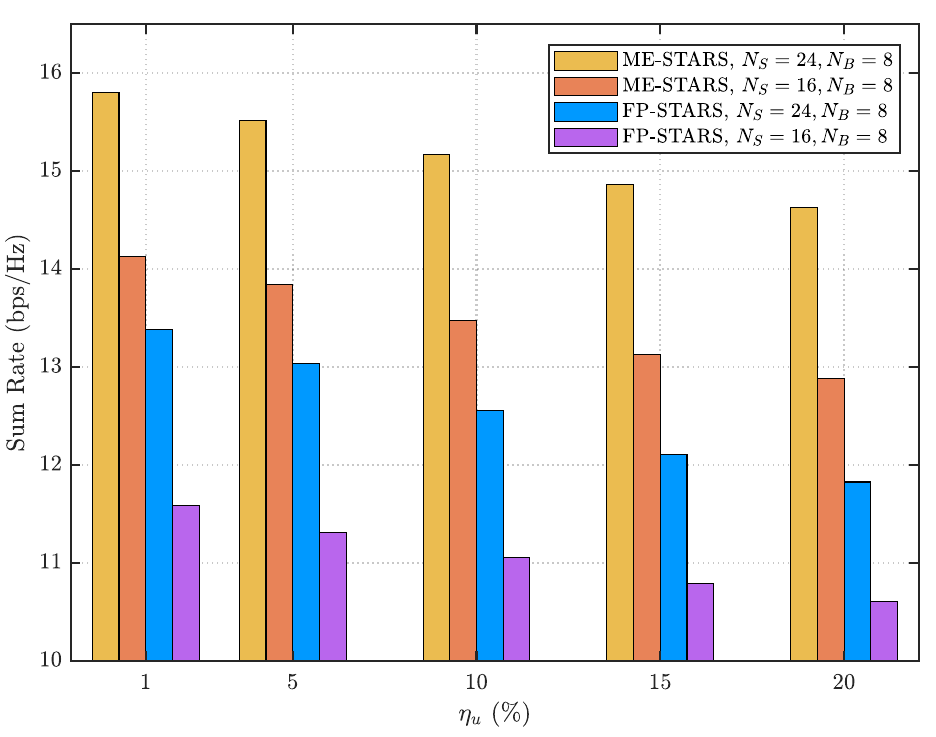}
	\caption{Sum rate versus $\eta_u$ for different $N_{\rm S}$.}
	\label{f7}
\end{figure} 

\begin{figure}[h]
	\centering
	\includegraphics [width=0.38\textwidth]{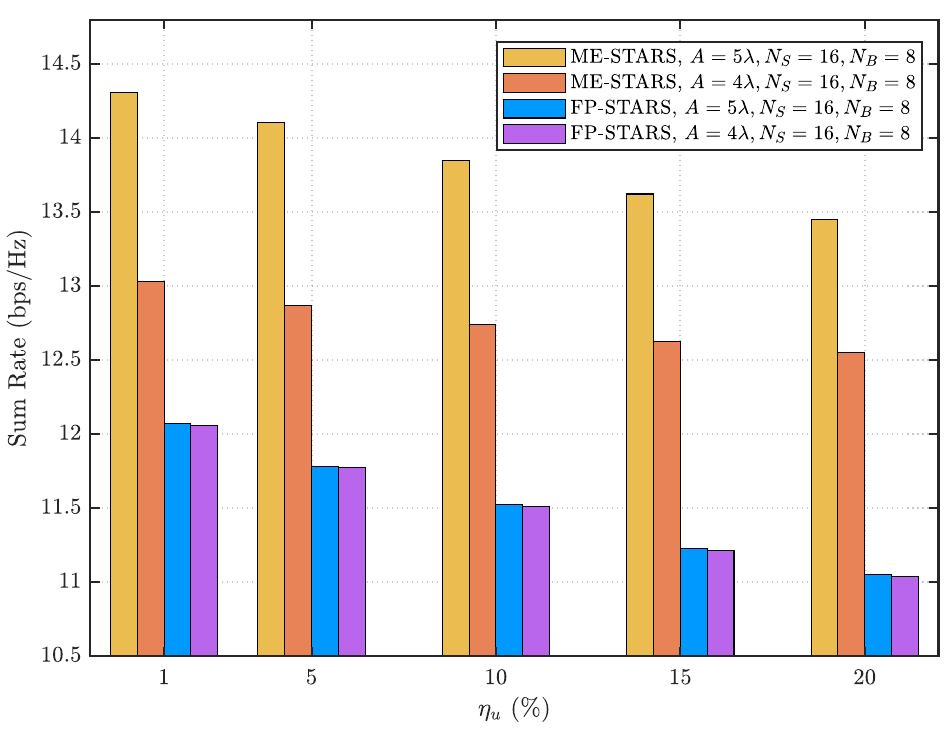}
	\caption{Sum rate versus $\eta_u$ for different movement-region sizes.}
	\label{f8}
\end{figure} 

\begin{figure}[h]
	\centering
	\includegraphics [width=0.38\textwidth]{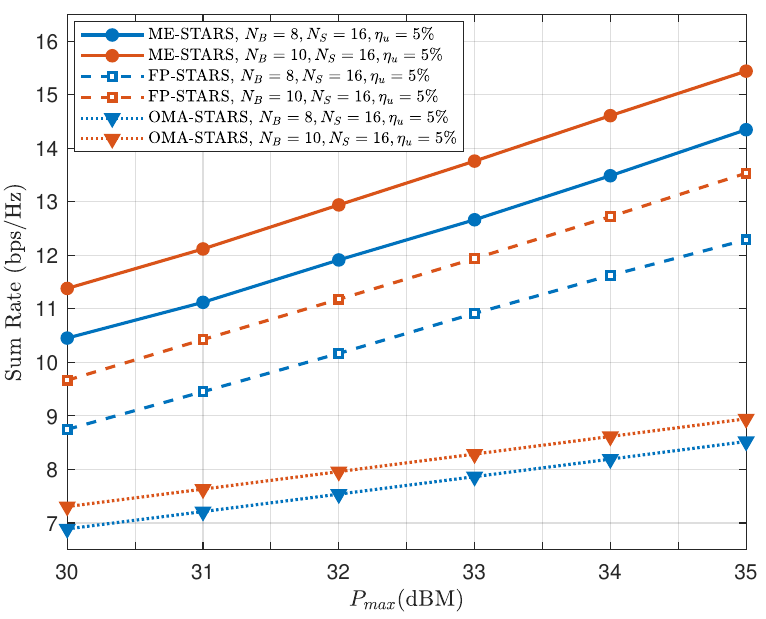}
	\caption{Sum rate versus $P_{\max}$ for different $N_{\rm B}$.}
	\label{f10}
\end{figure}

\begin{figure}[h]
	\centering
	\includegraphics [width=0.38\textwidth]{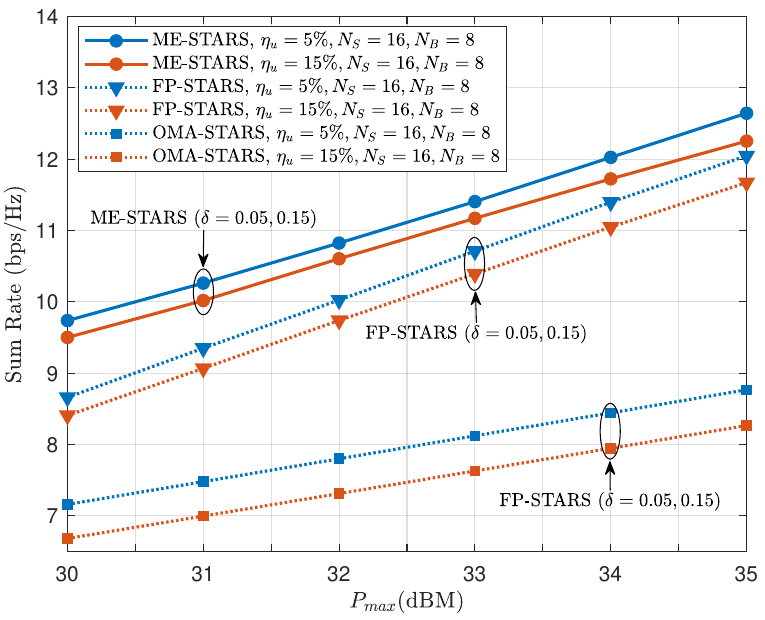}
	\caption{Sum rate versus $P_{\max}$ for different $\eta_u$.}
	\label{f9}
\end{figure} 

The convergence behavior of the proposed algorithm is illustrated in Figs.~\ref{f2} and \ref{f3} for the proposed ME-STARS and its fixed-position STARS (FP-STARS) counterpart, where the element positions of FP-STARS remain fixed throughout the optimization. It can be observed that the sum rate increases rapidly during the initial iterations and gradually approaches a stable value within a limited number of iterations, demonstrating the stable and fast convergence of the proposed iterative optimization framework. In Fig.~\ref{f2}, the converged sum rate decreases as the channel uncertainty level $\eta_u$ increases for both ME-STARS and FP-STARS, since larger CSI uncertainty limits the achievable performance under the robust transmission design. Nevertheless, ME-STARS consistently achieves a higher sum rate than FP-STARS for all considered uncertainty levels, demonstrating the performance gain provided by optimizing the element positions. Moreover, Fig.~\ref{f3} shows that increasing the number of STARS elements $N_{\rm S}$ improves the achievable sum rate for both schemes due to the increased spatial degrees of freedom. The additional gain achieved by ME-STARS over its fixed-position counterpart further demonstrates the benefit of element repositioning in establishing more favorable cascaded channel conditions.

Figs.~\ref{f4} and \ref{f5} illustrate the impact of the number of ME-STARS elements $N_{\rm S}$ on the achievable sum rate under different CSI uncertainty levels and numbers of BS antennas, respectively. In both
figures, the sum rate increases consistently with $N_{\rm S}$ for ME-STARS and FP-STARS, as a larger number of surface elements provides additional spatial degrees of freedom for controlling the cascaded BS--STARS--user channels. Fig.~\ref{f4} further shows that increasing the CSI uncertainty level $\eta_u$ degrades the achievable sum rate, whereas Fig.~\ref{f5} demonstrates that increasing the number of BS antennas improves the performance due to the enhanced transmit beamforming capability. More importantly, ME-STARS consistently outperforms its fixed-position counterpart over all considered values of $N_{\rm S}$, $\eta_u$, and $N_{\rm B}$, confirming the additional gain provided by optimizing the element positions.

Fig.~\ref{f6} shows the achievable sum rate versus the maximum transmit power $P_{\max}$ for different numbers of STARS elements. The sum rate increases steadily with $P_{\max}$ for all considered schemes, since a larger transmit-power budget enables the BS to deliver stronger useful signals to the users. A larger $N_{\rm S}$ further improves the performance by providing additional spatial degrees of freedom for shaping the cascaded BS--STARS--user channels. Moreover, ME-STARS consistently achieves a higher sum rate than its FP-STARS counterpart over the entire range of $P_{\max}$, demonstrating the additional gain provided by optimizing the element positions to obtain more favorable cascaded channel conditions.

Figs.~\ref{f7} and \ref{f8} illustrate the impact of CSI uncertainty $\eta_u$ on the achievable sum rate for different numbers of ME-STARS elements and movement-region sizes, respectively. The sum rate gradually decreases as $\eta_u$ increases, since larger channel uncertainty limits the achievable performance of the robust transmission design. Fig.~\ref{f7} shows that increasing $N_{\rm S}$ improves the sum rate for both ME-STARS and FP-STARS, while ME-STARS maintains a consistent performance advantage over its fixed-position counterpart. Furthermore, Fig.~\ref{f8} shows that enlarging the movement region provides an additional gain for ME-STARS by allowing the elements to explore a wider range of positions and establish more favorable cascaded channels. In contrast, the performance of FP-STARS remains almost unchanged with the movement-region size because its element positions are fixed.

Figs.~\ref{f9} and \ref{f10} show the achievable sum rate versus the maximum transmit power $P_{\max}$ for different numbers of BS antennas and CSI uncertainty levels, respectively. In addition to FP-STARS, we include an orthogonal multiple access STARS (OMA-STARS) counterpart, in which the users are served over orthogonal resources instead of employing RSMA. The sum rate increases with $P_{\max}$ for all considered schemes.
Fig.~\ref{f9} shows that increasing $N_{\rm B}$ improves the performance due to the enhanced transmit beamforming capability, whereas Fig.~\ref{f10} shows that a larger CSI uncertainty level $\eta_u$ reduces the achievable sum rate. More importantly, ME-STARS consistently outperforms both FP-STARS and OMA-STARS over the entire range of $P_{\max}$, highlighting the gains offered by element-position optimization and RSMA-based transmission.

\section{Conclusion}
This paper investigated robust transmission design for an ME-STARS assisted RSMA system under imperfect CSI, where movable STARS elements provide additional spatial degrees of freedom by adjusting their positions within a predefined region. A robust sum-rate maximization problem was formulated by jointly optimizing the transmit beamforming, common-rate allocation, reflection and transmission coefficients, and ME-STARS element positions subject to the transmit-power, user-rate, minimum inter-element spacing, and movement-region constraints. To solve the resulting non-convex problem, an iterative optimization framework was developed, with the element positions sequentially updated using an MM-based approach and quadratic surrogate functions constructed from the first- and second-order derivatives of the position-dependent channels. Simulation results demonstrated that the proposed ME-STARS design consistently outperforms the considered benchmark schemes, highlighting
the potential of element repositioning as an additional spatial design dimension for robust RSMA-based wireless communication systems.

\bibliographystyle{IEEEtran}
\bibliography{Ref}
\vskip -2\baselineskip plus -2fil

\end{document}